\documentclass[letterpaper,twocolumn,10pt]{article}
\usepackage{usenix2019_v3}

\usepackage{adjustbox}
\usepackage{amsmath,amssymb,amsfonts}
\usepackage{algorithmic}
\usepackage{balance}
\usepackage{enumitem}
\usepackage{etoolbox}
\usepackage{booktabs}
\usepackage[font={footnotesize}]{caption}
\usepackage{cite}
\usepackage{color, colortbl}
\usepackage{dblfloatfix}
\usepackage{enumitem}
\usepackage{graphicx}
\usepackage{cleveref} %
\usepackage{listings}
\usepackage{mdframed}
\usepackage{multirow}
\usepackage[numbers,sort]{natbib}
  \def\BibTeX{{\rm B\kern-.05em{\sc i\kern-.025em b}\kern-.08em
    T\kern-.1667em\lower.7ex\hbox{E}\kern-.125emX}}
\usepackage{pdfcomment}
\usepackage{relsize}
\usepackage{siunitx}
\usepackage{soul}
\usepackage{subcaption}
\usepackage{tabularx}
\usepackage{textcomp}
\usepackage{wasysym}
\usepackage{xcolor}
\usepackage{xspace}
\usepackage{float}

\newcommand{\head}[1]{\textnormal{\textbf{#1}}}

\def\Snospace~{\S{}}

\usepackage{tikz}
\usepackage{pgfplots}
\pgfplotsset{compat=1.13}
\usetikzlibrary{calc}
\usetikzlibrary{positioning,arrows,fit,matrix}
\usetikzlibrary{backgrounds}
\usetikzlibrary{patterns}
\usetikzlibrary{shapes.arrows} %
\usepgfplotslibrary{groupplots}
\usepgfplotslibrary{statistics}
\usepgfplotslibrary{fillbetween}

\tikzset{
	font=\footnotesize
}

\definecolor{cbone}  {HTML}{006BA4} %
\definecolor{cbtwo}  {HTML}{FF800E} %
\definecolor{cbthree}{HTML}{ABABAB} %
\definecolor{cbfour} {HTML}{595959} %
\definecolor{cbfive} {HTML}{5F9ED1} %
\definecolor{cbsix}  {HTML}{C85200} %
\definecolor{cbseven}{HTML}{898989} %
\definecolor{cbeight}{HTML}{A2C8EC} %
\definecolor{cbnine} {HTML}{FFBC79} %
\definecolor{cbten}  {HTML}{CFCFCF} %

\definecolor{applegreen}{rgb}{0.55, 0.71, 0.0}
\definecolor{goeblue}{RGB}{0,51,102}
\definecolor{tubsredSec}{cmyk}{0.0,1.00,0.6,0.6}
\definecolor{tubsredPrim}{cmyk}{0.1,1.0,0.8,0.0}

\iftrue
  \colorlet{primarycolor}{cbone}
  \colorlet{primaryshaded}{cbfive}
  \colorlet{secondarycolor}{cbtwo}
  \colorlet{secondaryshaded}{cbnine}

\else %

  \colorlet{primarycolor}{cyan!50!black}
  \colorlet{primaryshaded}{goeblue!50}
  \colorlet{secondarycolor}{tubsredSec!50}
  \colorlet{secondaryshaded}{tubsredPrim!50}
\fi

\definecolor{pro}{HTML}{60A7D2}
\definecolor{con}{HTML}{F18D2F}
\definecolor{literal}{HTML}{4F78A6}
\definecolor{select}{HTML}{105BA4}

\definecolor{stronglyagreecolor}  {HTML}{5F9ED1}
\definecolor{agreecolor}  {HTML}{A2C8EC}
\definecolor{neithercolor}  {HTML}{CFCFCF}
\definecolor{disagreecolor}  {HTML}{FFDAB9}
\definecolor{stronglydisagreecolor}  {HTML}{FF800E}
\colorlet{noanswercolor}{gray!20}

\pgfplotsset{
	coline1/.style={color=goeblue},
	coline2/.style={color=tubsredPrim!80!black, densely dashed},
	coline3/.style={color=red!50!black, densely dotted},
	coline4/.style={color=green!50!black, densely dashdotted},
	coline5/.style={color=purple, densely dashdotdotted},
	coline6/.style={color=gray!50!black, dotted},
	coline7/.style={color=blue!50!black, loosely dashdotdotted},
	coline8/.style={color=red!50!black, loosely dashdotted},
}

\newcommand{\resp}[1]{\pdfmargincomment[color=applegreen]{Responsible:~#1}}
\renewcommand{\resp}[1]{}

\newcommand{\az}{AndroZoo\xspace}
\newcommand{\gp}{GooglePlay\xspace}
\newcommand\CC{C\texttt{++}\xspace}%
\newcommand{\drebin}{\textsc{Drebin}\xspace}
\newcommand{\opseqs}{\textsc{Opseqs}\xspace}
\newcommand{\kitsune}{\textsc{Kitsune}\xspace}

\newcommand{\npapers}{\num{30}\xspace}  %

\renewcommand{\paragraph}[1]{{\vskip 4pt \noindent\textbf{#1.} }}

\mdfdefinestyle{MyFrame}{
	linecolor=black,
	backgroundcolor=gray!10,
	innertopmargin=5pt,
	innerbottommargin=10pt,
	innerrightmargin=5pt,
	innerleftmargin=5pt,
}
\newcommand{\pitfallenv}[4]{%
	\vspace{2mm}
	\noindent\begin{minipage}{\columnwidth}%
	\begin{mdframed}[style=MyFrame]
		\textbf{#1 -- #2.} #3
	\end{mdframed}%
	\vspace*{-14.35pt}%
		\includegraphics[width=1\columnwidth+0.0pt]{#4}%
\end{minipage}%
}

\mdfdefinestyle{HighlightFrame}{
	linecolor=black,
	backgroundcolor=white,
	innertopmargin=5pt,
	innerbottommargin=5pt,
	innerrightmargin=5pt,
	innerleftmargin=5pt,
}

\newcommand{\eg}{e.g.,\xspace} %

\newcommand{\perc}[1]{%
  \ifstrempty{#1}%
  {\,\si{\percent}}%
  {\SI{#1}{\percent}}%
}

\newcommand{\code}[1]{\textsmaller[0]{\texttt{#1}}\xspace}

\lstdefinestyle{mycppstyle}
{
	language=C++,
	keywordstyle=\bfseries,
	keywordstyle = [2]{\color{literal}},
	keywordstyle = [3]{\itshape},
	stringstyle=\color{literal},
	commentstyle=\color{gray},
	otherkeywords = {NULL, 50, 1, 00, wmemset, memmove, constexpr},
	morekeywords = [2]{NULL, 50, 1, 00},
	morekeywords = [3]{wmemset, memmove},
}

\newcommand{\hlc}[2][yellow]{{%
		\colorlet{foo}{#1}%
		\sethlcolor{foo}\hl{#2}}%
}

\lstnewenvironment{codelst}[1][]{%
	\lstset{%
		escapeinside={(*@}{@*)},
		basicstyle=\footnotesize\ttfamily,
		breaklines=true,
		xleftmargin=5pt,
		framexleftmargin=5pt,
		frame=single,
		numbers=left,
		numberstyle=\tiny,
		belowskip=0em,
		#1
	}%
}{}

\newcommand{\noemph}[1]{#1}
\newcommand{\samplingbias}        {\hyperref[pit:sampling-bias]         {\noemph{sampling bias}}\xspace}
\newcommand{\falsecausality}      {\hyperref[pit:spurious-correlations] {\noemph{spurious correlations}}\xspace}
\newcommand{\falsecausalities}    {\hyperref[pit:spurious-correlations] {\noemph{spurious correlations}}\xspace}

\newcommand{\datasnooping}        {\hyperref[pit:data-snooping]         {\noemph{data snooping}}\xspace}
\newcommand{\labelinaccuracy}     {\hyperref[pit:label-inaccuracy]      {\noemph{label inaccuracy}}\xspace}

\newcommand{\inapprmetrics}       {\hyperref[pit:inappropriate-metrics]{\noemph{inappropriate performance measures}}\xspace}
\newcommand{\apprbaseline}        {\hyperref[pit:inappropriate-baseline]{\noemph{appropriate baseline}}\xspace}
\newcommand{\flawedparamselection}{\hyperref[pit:flawed-param-selection]{\noemph{biased parameter selection}}\xspace}

\newcommand{\labonlyevaluations}  {\hyperref[pit:lab-only]              {\noemph{lab-only evaluations}}\xspace}
\newcommand{\adversarialsettings} {\hyperref[pit:attack]                {\noemph{inappropriate threat models}}\xspace}

\newcommand{\shortsamplingbias}        {\hyperref[pit:sampling-bias]         {P1}\xspace}
\newcommand{\shortlabelinaccuracy}     {\hyperref[pit:label-inaccuracy]      {P2}\xspace}
\newcommand{\shortdatasnooping}        {\hyperref[pit:data-snooping]         {P3}\xspace}
\newcommand{\shortfalsecausality}      {\hyperref[pit:spurious-correlations]       {P4}\xspace}
\newcommand{\shortflawedparamselection}{\hyperref[pit:flawed-param-selection]{P5}\xspace}
\newcommand{\shortinapprbaseline}      {\hyperref[pit:inappropriate-baseline]{P6}\xspace}
\newcommand{\shortinapprmetrics}      {\hyperref[pit:inappropriate-metrics]{P7}\xspace}
\newcommand{\shortbaseratefallacy}     {\hyperref[pit:base-rate]             {P8}\xspace}
\newcommand{\shortlabonlyevaluation}{\hyperref[pit:lab-only]                 {P9}\xspace}
\newcommand{\shortadversarialsetting}{{P10}\xspace}

\begin{document}

\date{}

\title{Dos and Don\textquotesingle ts of Machine Learning in Computer
  Security\thanks{To appear at {USENIX} Security Symposium 2022}\\}

\author{
{\rm 
    Daniel~Arp\footnotemark[1]~,~%
    Erwin~Quiring\footnotemark[2]~,~%
    Feargus~Pendlebury\footnotemark[3]~\footnotemark[4]~,~%
    Alexander~Warnecke\footnotemark[2]~,~%
    Fabio~Pierazzi\footnotemark[3]~,~%
  }
  \\
  {\rm
    Christian~Wressnegger\footnotemark[5]~,~%
    Lorenzo~Cavallaro\footnotemark[6]~,~%
    Konrad~Rieck\footnotemark[2]
  }\\
{\normalsize \footnotemark[1]~~Technische Universität Berlin}\\
{\normalsize \footnotemark[2]~~Technische Universität Braunschweig}\\
{\normalsize \footnotemark[3]~~King's College London},
{\normalsize \footnotemark[6]~~University College London}\\
{\normalsize \footnotemark[4]~~Royal Holloway, University of London and The Alan Turing Institute}\\
{\normalsize \footnotemark[5]~~KASTEL Security Research Labs and Karlsruhe Institute of Technology}\\
}
\maketitle

\begin{abstract}
  With the growing processing power of computing systems and the
  increasing availability of massive datasets, machine learning
  algorithms have led to major breakthroughs in many different areas. This
  development has influenced computer security, spawning a series
  of work on learning-based security systems, such as for malware
  detection, vulnerability discovery, and binary code
  analysis. Despite great potential, machine learning in
  security is prone to subtle pitfalls that undermine its  
  performance and render learning-based systems potentially unsuitable for  
  security tasks and practical deployment.

  In this paper, we look at this problem with critical eyes. First, we
  identify common pitfalls in the design, implementation, and
  evaluation of learning-based security systems. We conduct a study of
  \npapers~papers from top-tier security conferences within the past
  \num{10}~years, confirming that these pitfalls are widespread in the
  current security literature.  In an empirical analysis, we further
  demonstrate how individual pitfalls can lead to unrealistic
  performance and interpretations, obstructing the understanding of
  the security problem at hand.  As a remedy, we propose actionable
  recommendations to support researchers in avoiding or mitigating the
  pitfalls where possible. Furthermore, we identify open problems when
  applying machine learning in security and provide directions for
  further research.
\end{abstract}

\section{Introduction}

No day goes by without reading machine learning success stories.  The
widespread access to specialized computational resources and large
datasets, along with novel concepts and architectures for deep
learning, have paved the way for machine learning breakthroughs in
several areas, such as the translation of natural
languages~\citep{ChoMerBul+14, SutVinLe14, BahChoBen15} and the
recognition of image content~\citep{KriSutHin12, HeZhaRenSun16,
  SimZis15}.
This development has naturally influenced security research: although
mostly confined to specific applications in the
past~\citep[][]{Forrest96, WanParSto06, Fredrikson10}, machine
learning has now become one of the key enablers to studying and
addressing security-relevant problems at large in several application
domains, including intrusion detection \citep[][]{MirDoiEloSha18,
  DuLiZheSri+17}, malware analysis \citep[][]{JanBruVen11,
  MarOnwAndCriRosStr17}, vulnerability
discovery~\citep[][]{LiZouXuOu+18, YamMaiGasRie+15}, and binary
code analysis \citep{XuLiuFenYin+17, DinFunCha19, ShiSonMoa15}.

\begin{figure*}[t]
  \centering
  \includegraphics[width=\textwidth]{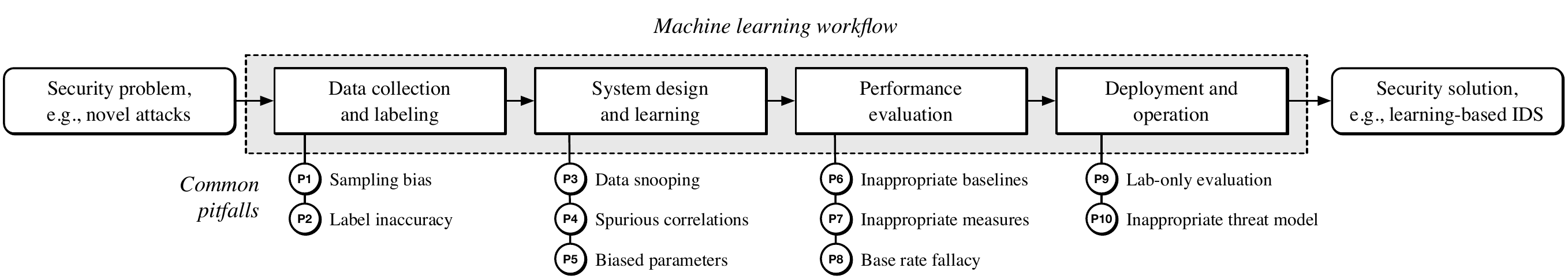}
  \caption{Common pitfalls of machine learning in computer security. }
  \label{fig:overview}
  \vspace{-0.5em}
\end{figure*}

Machine learning, however, has no clairvoyant abilities and
requires reasoning about statistical properties of data across a
fairly delicate workflow: incorrect assumptions and experimental biases
may cast doubts on this process to the extent that it becomes unclear
whether we can trust scientific discoveries made using learning
algorithms at all~\citep[][]{web:Feb19}. 
Attempts to identify such challenges and limitations in specific
security domains, such as network intrusion detection, started two decades
ago~\citep[][]{Axe00, TanMax02, sommer2010outside} and were extended
more recently to other domains, such as malware analysis and website
fingerprinting~\citep[][]{RosDieGriKrePaxPohBosSte12, packware:ndss20,
PenPieJor19,
  JuaAfrAcaDiaGre14}. Orthogonal to this line of work, however, we
argue that there exist \emph{generic pitfalls} related to machine
learning that affect all security domains and have received little
attention so far.

These pitfalls can lead to over-optimistic results and, even worse,
affect the entire machine learning workflow, weakening assumptions,
conclusions, and lessons learned. As a consequence, a false sense of
achievement is felt that hinders the adoption of research advances in
academia and industry. A sound scientific methodology is fundamental
to support intuitions and draw conclusions.  We argue that this need
is especially relevant in security, where processes are often
undermined by adversaries that actively aim to bypass analysis and
break systems.

In this paper, we identify ten~common---yet subtle---pitfalls that
pose a threat to validity and hinder interpretation of research
results. To support this claim, we analyze the prevalence of these
pitfalls in \npapers top-tier security papers from the
past decade that rely on machine learning for tackling different
problems.  To our surprise, each paper suffers from at least three
pitfalls; even worse, several pitfalls affect most of the papers,
which shows how endemic and subtle the problem is. Although the pitfalls
are widespread, it is perhaps more important to understand the extent
to which they weaken results and lead to over-optimistic conclusions. To this end, we
perform an impact analysis of the pitfalls in four different security
fields. The findings support our premise echoing the broader concerns
of the community.\\[-6pt]

\noindent In summary, we make the following contributions:

\begin{enumerate}%
\setlength{\itemsep}{3pt}

\item {\bf Pitfall Identification.} We identify ten~pitfalls as
  \emph{don'ts} of machine learning in security and propose
  \emph{dos} as actionable recommendations to support researchers in
  avoiding the pitfalls where possible. Furthermore, we identify open
  problems that cannot be mitigated easily and require further
  research effort
  (\autoref{sec:pitfalls}).

\item {\bf Prevalence Analysis.} We analyze the prevalence of the
  identified pitfalls in \npapers{}~representative top-tier security
  papers published in the past decade. Additionally, we perform a
  broad survey in which we obtain and evaluate the feedback of the
  authors of these papers regarding the identified pitfalls
  (\autoref{sec:prevalence}).

\item {\bf Impact Analysis.} In four different security domains, we
  experimentally analyze the extent to which such pitfalls introduce
  experimental bias, and how we can effectively overcome these
  problems by applying the proposed recommendations
  (\autoref{sec:casestudy}).

\end{enumerate}

\begin{mdframed}[style=HighlightFrame]
  \textbf{Remark.} This work should not be interpreted as a
  finger-pointing exercise. On the contrary, it is a reflective effort
  that shows how subtle pitfalls
  can have a negative impact on progress of security research, and how
  we---as a community---can mitigate them adequately.
 \end{mdframed}%

\section{Pitfalls in Machine Learning}
\label{sec:pitfalls}

Despite its great success, the application of machine learning in practice is often non-trivial and prone to
several pitfalls, ranging from obvious flaws to minor blemishes. Overlooking these
issues may result in experimental bias or incorrect conclusions,
especially in computer security. In this section, we present ten
common pitfalls that occur frequently in security research. Although
some of these pitfalls may seem obvious at first glance, they are
rooted in subtle deficiencies that are widespread in security
research---even in papers presented at top conferences
(see~\autoref{sec:prevalence} and~\autoref{sec:casestudy}).

We group these pitfalls with respect to the stages of a typical
machine learning workflow, as depicted in \autoref{fig:overview}.
For each pitfall, we provide a short description, discuss its impact
on the security domain, and present recommendations. Moreover, a
colored bar depicts the proportion of papers in our analysis that
suffer from the pitfall, with warmer colors indicating the presence of
the pitfall (see~\autoref{fig:survey-plot}).

\subsection{Data Collection and Labeling}

The design and development of learning-based systems usually starts
with the acquisition of a representative dataset. 
It is clear that conducting experiments using unrealistic data leads to the
misestimation of an approach's capabilities. The following two
pitfalls frequently induce this problem and thus require special
attention when developing learning-based systems in computer security.

\pitfallenv{P1}{Sampling Bias}%
{\label{pit:sampling-bias} The collected data does not sufficiently
represent the true data distribution of the underlying security
problem~\cite[]{CorMohRilRos08,YasMalHsu12,ChiFre18}.}%
{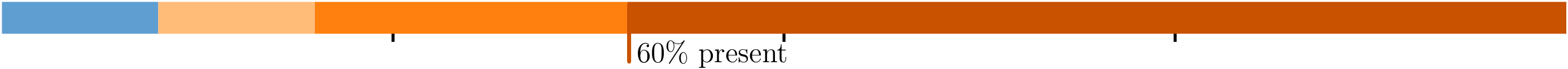}

\paragraph{Description} With a few rare exceptions, researchers
develop learning-based approaches without exact knowledge of the
true underlying distribution of the input space. Instead, they need to
rely on a dataset containing a fixed number of samples that aim to
resemble the actual distribution. While it is inevitable that some
bias exists in most cases, understanding the specific bias inherent to
a particular problem is crucial to limiting its impact in practice.
Drawing meaningful conclusions from the training data becomes
challenging, if the data does not effectively represent the input
space or even follows a different distribution.

\paragraph{Security implications} Sampling bias is highly relevant to
security, as the acquisition of data is particularly challenging and
often requires using multiple sources of varying quality. As an
example, for the collection of suitable datasets for Android malware
detection only a few public sources exist from which to obtain such
data~\citep{WeiLiRoy17,AllBisKleTra16}. As a result, it is common
practice to rely on synthetic data or to combine data from different
sources, both of which can introduce bias as we demonstrate
in~\autoref{sec:casestudy} with examples on 
state-of-the-art methods for intrusion and malware detection.

\paragraph{Recommendations} In many security applications, sampling
from the true distribution is extremely difficult and sometimes even
impossible.  Consequently, this bias can often only be mitigated but
not entirely removed.  In \autoref{sec:casestudy}, we show that, in
some cases, a reasonable strategy is to construct different estimates
of the true distribution and analyze them individually.  Further
strategies include the extension of the dataset with synthetic
data~\citep[\eg][]{ChaBowHalKeg02,HanWanMao05,WonGatStaMcD16} or the
use of transfer
learning~\citep[see][]{PalPomHinMit09,WeiKhoWan16,ZhuQiDuaXi21,ZhuXiSonZhu20}.
However, the mixing of data from incompatible sources should be
avoided, as it is a common cause of additional bias. In any case,
limitations of the used dataset should be openly discussed, allowing
other researchers to better understand the security implications of
potential sampling bias.

\pitfallenv{P2}{Label Inaccuracy}%
{\label{pit:label-inaccuracy} The ground-truth labels required for
classification tasks are inaccurate, unstable, or erroneous, affecting
the overall performance of a learning-based system \citep{LipWanSmo18,
ZhuShiYan+20}.}%
{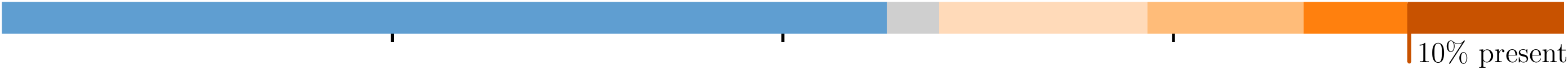}

\paragraph{Description} Many learning-based security
systems are built for classification tasks. To train these systems,
a ground-truth label is required for each observation. 
Unfortunately, this labeling is rarely perfect and
researchers must account for uncertainty and noise to prevent their
models from suffering from inherent bias.

\paragraph{Security implications} For many relevant security
problems, such as detecting network attacks or malware, reliable labels
are typically not available, resulting in a chicken-and-egg problem.
As a remedy, researchers often resort to heuristics, such as using
external sources that do not provide a reliable ground-truth. For
example, services like \emph{VirusTotal} are commonly used for
acquiring label information for malware but these are not always 
consistent~\citep{ZhuShiYan+20}.
Additionally, changes in adversary behavior may alter the ratio between
different classes over
time~\citep{MilKanTscAfrBacFai16,packware:ndss20,ZhuShiYan+20}, 
introducing a bias known as \emph{label shift}~\cite{LipWanSmo18}. A system
that cannot adapt to these changes will experience performance decay 
once deployed.

\paragraph{Recommendations} Generally, labels should be verified
whenever possible, for instance, by manually investigating false
positives or a random sample~\citep[\eg][]{StoCis18}. If \emph{noisy
  labels} cannot be ruled out, their impact on the learning model can
be reduced by (i)~using robust models or loss functions, (ii)~actively
modeling label noise in the learning process, or (iii)~cleansing noisy
labels in the training data~\citep[see][]{FreVer14, HurTanDas17,
  LiaGuoLuo21}.  To demonstrate the applicability of such approaches,
we empirically apply a cleansing approach in
\autoref{sec:appendix_pitfalls}.
Note that instances with uncertain labels must not be removed from the 
test data. This represents a variation of sampling bias
(\shortsamplingbias) and data snooping (\shortdatasnooping), a pitfall
we discuss in detail in \autoref{subsec:system_design}.
Furthermore, as labels may change over time, it is necessary to take precautions 
against \emph{label shift}~\cite{LipWanSmo18}, such as by delaying
labeling until a stable ground-truth is
available~\citep[see][]{ZhuShiYan+20}.

\subsection{System Design and Learning}
\label{subsec:system_design}

Once enough data has been collected, a learning-based security system
can be trained.  This process ranges from data preprocessing to
extracting meaningful features and building an effective learning
model. Unfortunately, flaws and
weak spots can be introduced at each of these steps.

\pitfallenv{P3}{Data Snooping}%
{\label{pit:data-snooping} A learning model is trained with data
that is typically not available in practice.  Data snooping can occur in
many ways, some of which are very subtle and hard to
identify~\citep{YasMalHsu12}.}%
{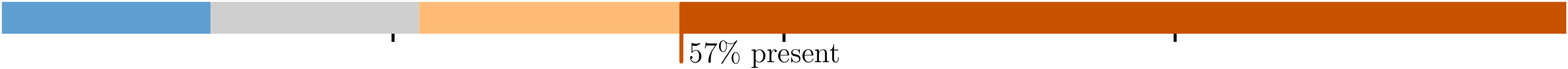}

\paragraph{Description} It is common practice to split collected
data into separate training and test sets prior to generating a
learning model. Although splitting the data seems straightforward,
there are many subtle ways in which test data or other background
information that is not usually available can affect the training
process, leading to data snooping. While a detailed list of data
snooping examples is provided in the appendix (see
\autoref{tab:appendix-snooping}), we broadly distinguish between three
types of data snooping: \emph{test}, \emph{temporal}, and
\emph{selective snooping.}

Test snooping occurs when the test set is used for experiments before
the final evaluation. This includes preparatory work to identify
useful features, parameters, and learning algorithms.  Temporal
snooping occurs if time dependencies within the data are ignored. This
is a common pitfall, as the underlying distributions in many
security-related problems are under continuous
change~\citep[\eg][]{MagRobKru09,PenPieJor19}. Finally, selective
snooping describes the cleansing of data based on information not
available in practice. An example is the removal of outliers based on
statistics of the complete dataset (i.e., training and test) that
are usually not available at training time.

\paragraph{Security implications} In security, data distributions are often 
non-stationary and continuously changing due to
new attacks or technologies. Because of this, snooping on data from 
the future or from external data sources is a prevalent pitfall that 
leads to over-optimistic results. 
For instance, several researchers have identified
temporal snooping in learning-based malware detection 
systems~\mbox{\citep[\eg][]{AllTaegBisKleTra15, AndSloBos17,
    PenPieJor19}}.  In all these cases, the capabilities of the
methods are overestimated due to mixing samples from past and
present. Similarly, there are incidents of test and selective
snooping in security research that lead to unintentionally biased results (see
\autoref{sec:prevalence}).

\paragraph{Recommendations} While it seems obvious that training,
validation, and test data should be strictly separated, this data
isolation is often unintentionally violated during the preprocessing
stages. For example, we observe that it is a common mistake to compute
tf-idf weights or neural embeddings over the entire dataset (see
\autoref{sec:prevalence}). To avoid this problem, test data should be
split early during data collection and stored separately until the
final evaluation. Furthermore, temporal dependencies within the data
should be considered when creating the dataset
splits~\citep{MagRobKru09,AllTaegBisKleTra15,PenPieJor19}.  Other
types of data snooping, however, are challenging to address.  For
instance, as the characteristics of publicly available datasets are
increasingly exposed, methods developed using this data implicitly
leverage knowledge from the test
data~\mbox{\citep[see][]{McH00,YasMalHsu12}}.  Consequently,
experiments on well-known datasets should be complemented with
experiments on more recent data from the considered application
domain.

\pitfallenv{P4}{Spurious Correlations}%
{\label{pit:spurious-correlations} Artifacts unrelated to the security
  problem create shortcut patterns for separating classes.
  Consequently, the learning model adapts to these artifacts instead
  of solving the actual task.}%
{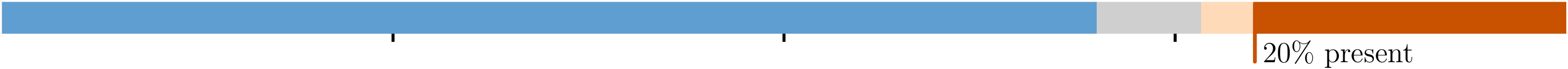}

\paragraph{Description} Spurious correlations result from artifacts
that correlate with the task to solve but are not actually related to
it, leading to false associations. Consider the example of a network
intrusion detection system, where a large fraction of the attacks in
the dataset originate from a certain network region. The model may
learn to detect a specific IP range instead of generic attack
patterns. Note that while sampling bias is a common reason for
spurious correlations, these can also result from other factors, as we
discuss in more detail in Appendix~\ref{sec:appendix_pitfalls}.

\paragraph{Security implications} Machine learning is typically
applied as a black box in security. As a result, \falsecausality often
remain unidentified.  
These correlations pose a
problem once results are interpreted and used for drawing general
conclusions. Without knowledge of \falsecausality, there is a high
risk of overestimating the capabilities of an approach and misjudging
its practical limitations.
As an example, \autoref{subsec:vulndiscovery} reports our analysis on
a vulnerability discovery system indicating the presence of
notable \falsecausality in the underlying data.

\paragraph{Recommendations} To gain a better view of the capabilities
of a learning-based systems, we generally recommend applying
explanation techniques for machine
learning~\mbox{\citep[see][]{GuoMuXuSu+18, LapWaeBinMonSamMue19,
    WarArpWres+20}}. Despite some
limitations~\citep[\eg][]{KinHooAde+19, HooErhKin+19, TomHarCha+20},
these techniques can reveal spurious correlations and allow a
practitioner to assess their impact on the system's capabilities. As
an example, we show for different security-related problems how
explainable learning can help to identify this issue in
\autoref{sec:casestudy}. Note that spurious correlations in one
setting may be considered a valid signal in another, depending on the
objective of the learning-based system.  Consequently, we recommend 
clearly defining this objective in advance and validating whether
correlations learned by the system comply with this goal. For example,
a robust malware detection system should pick up features related to
malicious activity rather than other unrelated information present in
the data.

\pitfallenv{P5}{Biased Parameter Selection}%
{\label{pit:flawed-param-selection} The final parameters of a
learning-based method are not entirely fixed at training time. Instead,
they indirectly depend on the test set.}%
{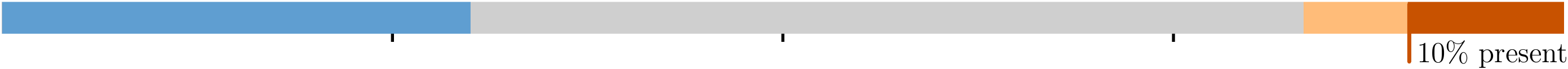}

\paragraph{Description} Throughout the learning procedure, it is
common practice to generate different models by varying
hyperparameters. The best-performing model is picked and its 
performance on the test set is presented.  While this 
setup is generally sound, it can still suffer
from a biased parameter selection. 
For example, over-optimistic results can be easily produced by tuning
hyperparameters or calibrating thresholds on the test data instead of
the training  data.

\paragraph{Security implications} A security system whose parameters
have not been fully calibrated at training time can perform very
differently in a realistic setting. While the detection threshold for
a network intrusion detection system may be chosen using a ROC curve
obtained on the test set, it can be hard to select the same
operational point in practice due the diversity of real-world
traffic~\citep{sommer2010outside}. This may lead to decreased
performance of the system in comparison to the original experimental
setting.  Note that this pitfall is related to
\datasnooping~(\shortdatasnooping), but should be considered
explicitly as it can easily lead to inflated results.

\paragraph{Recommendations} This pitfall constitutes a special case of
data snooping and thus the same countermeasures apply. However, in
practice fixing a biased parameter selection can often be easily
achieved by using a separate \emph{validation set} for model selection
and parameter~tuning. In contrast to general data snooping, which is
often challenging to mitigate, strict data isolation is already
sufficient to rule out problems when determining hyperparameters
and thresholds.

\subsection{Performance Evaluation}

The next stage in a typical machine-learning workflow is the
evaluation of the system's performance. In the following, we show how
different pitfalls can lead to unfair comparisons and biased results
in the evaluation of such systems.

\pitfallenv{P6}{Inappropriate Baseline}%
{\label{pit:inappropriate-baseline} The evaluation is conducted
  without, or with limited, baseline methods. As a result, it is
  impossible to demonstrate improvements against the state of the art
  and other security mechanisms.}%
{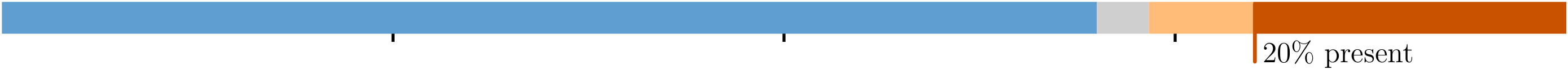}

\paragraph{Description} To show to what extent a novel method improves
the state of the art, it is vital to compare it with previously
proposed methods. 
When choosing baselines, it is important to remember that there exists
no universal learning algorithm that outperforms all other approaches
in general~\citep{Wol96}.
Consequently, providing only results for the proposed approach or a
comparison with mostly identical learning models, does not give enough
context to assess its impact.

\paragraph{Security implications} An overly complex learning method
increases the chances of overfitting, and also the runtime overhead,
the attack surface, and the time and costs for deployment. To show
that machine learning techniques provide significant improvements
compared to traditional methods, it is thus essential to compare these
systems side by side.

\paragraph{Recommendations} Instead of focusing solely on complex
models for comparison, simple models should also be considered
throughout the evaluation. These methods are easier to explain, less
computationally demanding, and have proven to be effective and
scalable in practice. In \autoref{sec:casestudy}, we demonstrate how
using well-understood, simple models as a baseline can expose
unnecessarily complex learning models. Similarly, we show that
automated machine learning \emph{(AutoML)}
frameworks~\citep[\eg][]{FeuKleEgg15,JinSonHu19} can help
finding proper baselines.
While these automated methods can certainly not replace experienced
data analysts, they can be used to set the lower bar the proposed approach
should aim for. Finally, it is critical to check whether non-learning
approaches are also suitable for the application scenario. For
example, for intrusion and malware detection, there exist a wide range
of methods using other detection
strategies~\citep[\eg][]{Pax98,Roe99,EncOngMcD09}.

\begin{figure}[t!]
	\centering
	\includegraphics{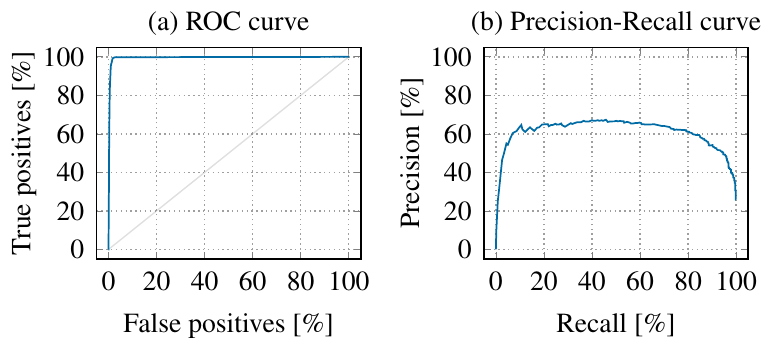}
	\vspace{-10px}
	\caption{ROC and precision-recall curve as two performance measures
	for the same scores, created on an artificial dataset with an
	imbalanced class ratio.
	Only the precision-recall curve conveys the true performance.}
	\label{fig:pitfall_metrics}
\end{figure}

\pitfallenv{P7}{Inappropriate Performance Measures}%
{\label{pit:inappropriate-metrics} The chosen performance measures do
not account for the constraints of the application scenario, such as
imbalanced data or the need to keep a low false-positive rate.}%
{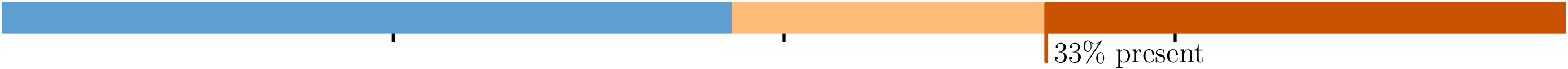}

\paragraph{Description} A wide range of performance measures are
available and not all of them are suitable in the context of security.
For example, when evaluating a detection system, it is typically
insufficient to report just a single performance value, such as the
accuracy, because true-positive and false-positive decisions are not
observable.  However, even more advanced measures, such as ROC curves,
may obscure experimental results in some application settings.
\autoref{fig:pitfall_metrics} shows an ROC curve and a
precision-recall curve on an imbalanced dataset (class
ratio~\num{1}:\num{100}).  Given the ROC curve alone, the performance
appears excellent, yet the low precision reveals the true performance
of the classifier, which would be impractical for many security
applications.

Furthermore, various security-related problems deal with more than two
classes, requiring \emph{multi-class metrics}. This setting can
introduce further subtle pitfalls.  Common strategies, such as
\emph{macro-averaging} or \emph{micro-averaging} are known to
overestimate and underestimate small classes~\citep{For04}.

\paragraph{Security implications} Inappropriate performance measures are a
long-standing problem in security research, particularly in detection
tasks.  While true and false positives, for instance, provide a more 
detailed picture of a system's performance, they can also disguise the 
actual precision when the prevalence of attacks is low. 

\paragraph{Recommendations} 
The choice of performance measures in machine learning is highly
application-specific. Hence, we refrain from providing general
guidelines.  Instead, we recommend considering the practical
deployment of a learning-based system and identifing measures that help
a practitioner assess its performance. Note that these measures
typically differ from standard metrics, such as the accuracy or error,
by being more aligned with day-to-day operation of the system.  To
give the reader an intuition, in \autoref{subsec:androidmalware}, we
show how different performance measures for an Android malware
detector lead to contradicting interpretations of its performance.

\pitfallenv{P8}{Base Rate Fallacy}%
{\label{pit:base-rate} A large class imbalance is ignored when
interpreting the performance measures leading to an overestimation 
of performance.}%
{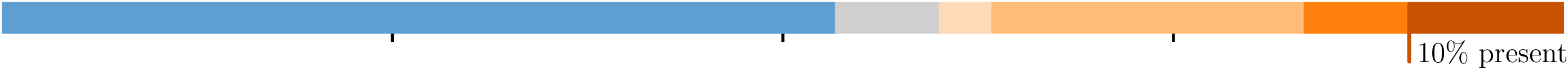}

\paragraph{Description} Class imbalance can easily lead to a
misinterpretation of performance if the base rate of the negative
class is not considered. If this class is predominant, even a very low
false-positive rate can result in surprisingly high numbers of false
positives.  Note the difference to the previous pitfall: while
\shortinapprmetrics refers to the inappropriate \emph{description} of
performance, the base-rate fallacy is about the misleading
\emph{interpretation} of results. This special case is easily
overlooked in practice~(see~\autoref{sec:prevalence}).
Consider the example in \autoref{fig:pitfall_metrics} where
\perc{99}~true positives are possible at \perc{1}~false
positives. Yet, if we consider the class ratio of
\mbox{\num{1}:\num{100}}, this actually corresponds to \num{100}~false
positives for every \num{99}~true positives.

\paragraph{Security implications}
The base rate fallacy is relevant in a variety of security problems,
such as intrusion detection and website fingerprinting
\cite[\eg][]{Axe00,JuaAfrAcaDiaGre14,PanLanZinHen16}.
As a result, it is challenging to realistically quantify the security 
and privacy threat posed by attackers. Similarly, the probability of
installing malware is usually much lower than is considered in
experiments on malware detection~\citep{PenPieJor19}.

\paragraph{Recommendations} Several problems in security revolve
around detecting rare events, such as threats and attacks. For these
problems, we advocate the use of \emph{precision} and \emph{recall} as
well as related measures, such as precision-recall curves. In contrast
to other measures, these functions account for class imbalance and
thus resemble reliable performance indicators for detection tasks
focusing on a minority class~\citep{SokLap09,
  davis2006relationship}. However, note that precision and recall can
be misleading if the prevalence of the minority class is inflated, for
example, due to sampling bias~\cite{PenPieJor19}.  In these cases,
other measures like \emph{Matthews Correlation Coefficient (MCC)} are
more suitable to assess the classifier's performance~\citep{ChiJur20}
(see~\autoref{sec:casestudy}).  In addition, ROC curves and their AUC
values are useful measures for comparing detection and classification
approaches. To put more focus on practical constraints, we recommend
considering the curves only up to tractable false-positive rates and 
to compute bounded AUC values. Finally, we also recommend discussing
false positives in relation to the base rate of the negative class,
which enables the reader to get an understanding of the workload
induced by false-positive decisions.

\subsection{Deployment and Operation}

In the last stage of a typical machine-learning workflow, the
developed system is deployed to tackle the underlying security problem
in practice. 

\pitfallenv{P9}{Lab-Only Evaluation}{\label{pit:lab-only} A
learning-based system is solely evaluated in a laboratory setting,
without discussing its practical limitations.}%
{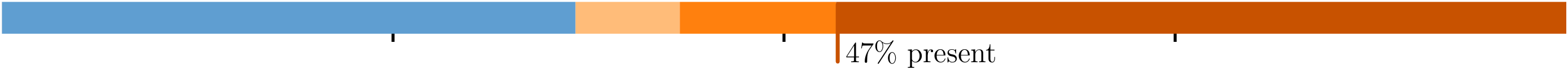}

\paragraph{Description} As in all empirical disciplines, it is common
to perform experiments under certain assumptions to demonstrate a
method's efficacy. While performing controlled experiments is a
legitimate way to examine specific aspects of an approach, it should
be evaluated in a realistic setting whenever possible to transparently
assess its capabilities and showcase the open challenges that will
foster further research.

\paragraph{Security implications} Many learning-based systems in
security are evaluated solely in laboratory settings, overstating
their practical impact. A common example are detection methods
evaluated only in a \emph{closed-world setting} with limited diversity
and no consideration of
non-stationarity~\citep{JorShaDas17,Barbero2020}. For example, a large
number of website fingerprinting attacks are evaluated only in
closed-world settings spanning a limited time
period~\citep{JuaAfrAcaDiaGre14}. Similarly, several learning-based
malware detection systems have been insufficiently examined under
realistic settings~\citep[see][]{Allix:Empirical, PenPieJor19}.

\paragraph{Recommendations} It is essential to move away from a
\textit{laboratory setting} and approximate a \textit{real-world
  setting} as accurately as possible.  For example, temporal and
spatial relations of the data should be considered to account for the
typical dynamics encountered in the wild~\citep[see][]{PenPieJor19}.
Similarly, runtime and storage constraints should be analyzed under
practical
conditions~\citep[see][]{RosDieGriKrePaxPohBosSte12,KouHeiAndBosGiu19,Barbero2020}.
Ideally, the proposed system should be deployed to uncover problems
that are not observable in a lab-only environment, such as the
diversity of real-world network
traffic~\citep[see][]{sommer2010outside}---although this is not 
always possible due to ethical and privacy constraints.

\pitfallenv{P10}{Inappropriate Threat Model}{\label{pit:attack} The
security of machine learning is not considered, exposing the system to a
variety of attacks, such as poisoning and evasion attacks.}%
{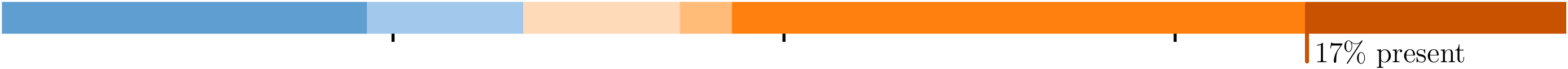}

\paragraph{Description} Learning-based security systems operate in a
hostile environment, which should be considered when designing
these systems. Prior work in adversarial learning has revealed a
considerable attack surface introduced by machine learning itself, at
all stages of the workflow~\mbox{\citep[see][]{biggio2018wild, PapMcSin+18}}.
Their broad attack surface makes these algorithms vulnerable to
various types of attacks, such as adversarial preprocessing, poisoning, 
and evasion~\citep[\eg][]{BigNelLas11, BigCorMai+13, CarWag17, 
pierazzi2020problemspace, QuiKleArp20}.

\paragraph{Security implications} Including adversarial influence in
the threat model and evaluation is often vital, as systems prone to
attacks are not guaranteed to output trustworthy and meaningful
results. Aside from traditional security issues, it is therefore
essential to also consider machine learning-related attacks.
For instance, an attacker may more easily evade a model that relies on
only a few features than a properly regularized model that has been
designed with security considerations in mind~\citep{DemMelBig17},
although one should also consider domain-specific
implications~\citep{pierazzi2020problemspace}.  Furthermore,
\emph{semantic gaps} in the workflow of machine learning may create
blind spots for attacks. For example, imprecise parsing and feature
extraction may enable an adversary to hide malicious
content~\citep{RndLas14}.

\paragraph{Recommendations} In most fields of security where
learning-based systems are used, we operate in an \emph{adversarial
  environment}. Hence, threat models should be defined precisely and
systems evaluated with respect to them. In most cases, it is necessary
to assume an \emph{adaptive adversary} that specifically targets the
proposed systems and will search for and exploit weaknesses for
evasion or manipulation.  Similarly, it is crucial to consider all
stages of the machine learning workflow and investigate possible
vulnerabilities \mbox{\citep[see][]{biggio2018wild, PapMcSin+18,
    CarAthPapBreRauTsiGooMadKur19, privacyml2020overview}}.  For this
analysis, we recommend focusing on white-box attacks where possible,
following Kerckhoff's principle~\cite{kerckhoffs1883cryptographie} and
security best practices.  Ultimately, we like to stress that an
evaluation of adversarial aspects is not an add-on but rather a 
mandatory component in security~research.

\newcommand{\present}{\emph{present}\xspace}
\newcommand{\ppresent}{\emph{partly present}\xspace}
\newcommand{\dpresent}{\emph{discussed}\xspace}
\newcommand{\notpresent}{\emph{not present}\xspace}
\newcommand{\unclear}{\emph{unclear from text}\xspace}
\newcommand{\nah}{\emph{does not apply}\xspace}

\begin{figure*}[t]
    \centering
    \noindent
    \includegraphics[width=\textwidth]{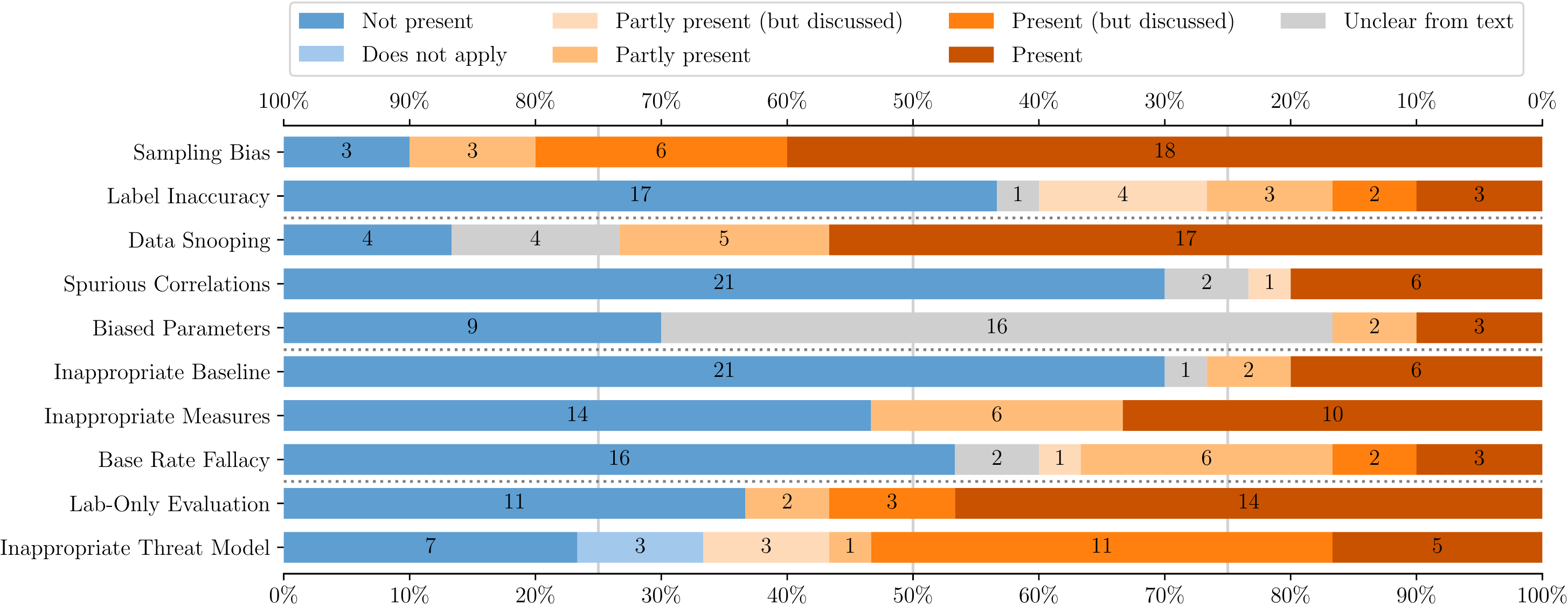}
    \vspace*{-6mm}
    \caption{Stacked bar chart showing the pitfalls suffered by each of
the \npapers~papers analyzed. The colors of each bar show the
degree to which a pitfall was present, and the width shows the
proportion of papers in that group. The number at the center of each
bar shows the cardinality of each group.
}
    \label{fig:survey-plot}
    \vspace*{-4mm}
\end{figure*}

\section{Prevalence Analysis\resp{Feargus (+Daniel)}}
\label{sec:prevalence}

Once we understand the pitfalls faced by learning-based security
systems, it becomes necessary to assess their prevalence and
investigate their impact on scientific advances.  To this end, we
conduct a study on \npapers papers published in the last ten years at
ACM~CCS, IEEE~S\&P, USENIX Security, and NDSS, the top-4 conferences
for security-related research in our community.
The papers have been selected as representative examples for our
study, as they address a large variety of security topics and
successfully apply machine learning to the corresponding research
problems.

In particular, our selection of top-tier papers covers the following
topics:
malware detection~\mbox{\citep{
    ArpSprHueGasRie14,
    MarOnwAndCriRosStr17,
    PenPieJor19,
    CurLivSei11,
    xi2019deepintent,
    SrnLas13}}; %
network intrusion detection~\mbox{\citep{
    DuLiZheSri+17,
    shen2018tiresias,
    MirDoiEloSha18,
    ShuYaoRam15}}; %
vulnerability discovery~\mbox{\citep{
    DinFunCha19,
	FisXiaKaoSta19,
    FisBoeXiaStrAcaBacFah17,
    LiZouXuOu+18}}; %
website fingerprinting attacks~\mbox{\citep{
	ShuKanHasMelMitOreYar19,
    RimPreJuaGoeJoo18,
    DyeCouRisShr12,
    PanLanZinHen16}}; %
social network abuse~\mbox{\citep{
	NilLabSedZan17,
    boshmaf2015integro,
    song2015crowdtarget}}; %
binary code analysis~\mbox{\citep{
	ShiSonMoa15,
	ChuSheSax+17,
    BaoBurWooTurBru14}}; %
code attribution~\citep{
    CalHarLiuNar+15,
    AbuAbuMoh+18}; %
steganography~\citep{
	barradas2018covert}; %
online scams~\citep{
    kharraz2018surveylance}; %
game bots~\citep{
    LeeWooKim+16}; %
and %
ad blocking~\citep{
	iqbal2020adgraph}%
.

\paragraph{Review process} Each paper is assigned two independent
reviewers who assess the article and identify instances of the
described pitfalls.  The pool of reviewers consists of six researchers
who have all previously published work on the topic of machine
learning and security in at least one of the considered security
conferences. Reviewers do \emph{not} consider any material presented
outside the papers under analysis (aside from appendices and
associated artifacts, such as datasets or source code).
Once both reviewers have completed their assignments, they discuss the
paper in the presence of a third reviewer that may resolve any disputes.
In case of uncertainty, the authors are given the benefit
of the doubt (\eg~in case of a dispute between \ppresent and \present,
we assign \ppresent).

Throughout the process, all reviewers meet regularly in order to discuss
their findings and ensure consistency between the pitfalls' criteria.
Moreover, these meetings have been used to refine the definitions and
scope of pitfalls based on the reviewers' experience. Following any
adaptation of the criteria, all completed reviews have been re-evaluated
by the original reviewers---this occurred twice during our analysis.
While cumbersome, this adaptive process of incorporating
reviewer feedback ensures that the pitfalls are comprehensive in
describing the core issues across the state of the art.
We note that the inter-rater reliability of reviews prior to dispute
resolution is $\alpha = 0.832$ using Krippendorff's alpha, where $\alpha
> 0.800$ indicates confidently reliable ratings~\cite{Kri18}.

\paragraph{Assessment criteria} For each paper, pitfalls are coarsely
classified as either \present, \notpresent, \unclear, or \nah.
A pitfall may be wholly present throughout the experiments
without remediation \emph{(present)}, or it may not \emph{(not present)}. If the
authors have corrected any bias or have narrowed down their claims to
accommodate the pitfall, this is also counted as \notpresent.
Additionally, we introduce \ppresent as a category to account for
experiments that do suffer from a pitfall, but where the impact has been
partially addressed.
If a pitfall is \present or \ppresent but acknowledged in the text, we
moderate the classification as \dpresent. If the reviewers are
unable to rule out the presence of a pitfall due to missing information,
we mark the publication as \unclear.
Finally, in the special case of~\shortadversarialsetting, if the pitfall \nah
to a paper's setting, this is considered as a separate category.

\paragraph{Observations}
The aggregated results from the prevalence analysis are shown in
\autoref{fig:survey-plot}. A bar's color indicates the degree to
which a pitfall is present, and its width shows the proportion of papers
with that classification. The number of affected papers is
noted at the center of the~bars.
The most prevalent pitfalls are \samplingbias~(\shortsamplingbias) and
\datasnooping~(\shortdatasnooping), which are at least partly present in 
\perc{90} and \perc{73} of the papers, respectively.
In more than \perc{50}~of the papers, we identify
\adversarialsettings~(\shortadversarialsetting),
\labonlyevaluations~(\shortlabonlyevaluation), and
\inapprmetrics~(\shortinapprmetrics) as at least partly present.
\textit{Every} paper is affected by at least three pitfalls,
underlining the pervasiveness of such issues in recent computer security
research.
In particular, we find that dataset collection is still very
challenging: some of the most realistic and expansive open datasets we
have developed as a community are still imperfect
(see~\autoref{subsec:androidmalware}).

Moreover, the presence of some pitfalls is more likely to be
\emph{unclear from the text} than others. We observe
this for \flawedparamselection~(\shortflawedparamselection) when no
description of the hyperparameters or tuning procedure is given; for
\falsecausality (\shortfalsecausality) when there is no attempt to
explain a model's decisions; and for~\datasnooping
(\shortdatasnooping) when the dataset splitting or normalization
procedure is not explicitly described in the text.
These issues also indicate that experimental settings are more difficult
to reproduce due to a lack of information.

\paragraph{Feedback from authors} To foster a discussion within our
community, we have contacted the authors of the selected papers and
collected feedback on our findings. We conducted a
survey with \num{135}~authors for whom contact information has been
available.  To protect the authors' privacy and encourage an open
discussion, all responses have been anonymized.

The survey consists of a series of general and specific questions on
the identified pitfalls. First, we ask the authors whether they have
read our work and consider it helpful for the community. Second, for
each pitfall, we collect feedback on whether they agree that (a) their
publication might be affected, (b) the pitfall frequently occurs in security
papers, and (c) it is easy to avoid in most cases.  To quantitatively
assess the responses, we use a five-point Likert scale for each
question that ranges from \emph{strongly disagree} to \emph{strongly
agree}. Additionally, we provide an option of \emph{prefer not to
answer} and allow the authors to omit questions.

We have received feedback from \num{49}~authors, yielding a response
rate of \perc{36}. These authors correspond to \num{13}~of the
\num{30}~selected papers and thus represent \perc{43}~of the considered
research. Regarding the general questions, \num{46}~(\perc{95}) of the
authors have read our paper and \num{48}~(\perc{98}) agree that it helps
to raise awareness for the identified pitfalls.
For the specific pitfall questions, the overall agreement between the
authors and our findings is~\perc{63} on average, varying depending on
the security area and pitfall. All authors agree that their paper may
suffer from at least one of the pitfalls. On average, they indicate that
\num{2.77}~pitfalls are present in their work with a standard deviation
of~\num{1.53} and covering all ten pitfalls.

When assessing the pitfalls in general, the authors especially agree
that lab-only evaluations~(\perc{92}), the base rate fallacy
(\perc{77}), inappropriate performance measures~(\perc{69}), and
sampling bias~(\perc{69}) frequently occur in security papers. Moreover,
they state that inappropriate performance measures~(\perc{62}),
inappropriate parameter selection~(\perc{62}), and the base rate
fallacy~(\perc{46}) can be easily avoided in practice, while the other
pitfalls require more effort. We provide further information on the
survey in Appendix~\ref{sec:appendix-survey}.

In summary, we derive three central observations from this survey.
First, most authors agree that there is a lack of awareness for the
identified pitfalls in our community. Second, they confirm that the
pitfalls are widespread in security literature and that there is a need
for mitigating them. Third, a consistent understanding of the identified
pitfalls is still lacking. As an example, several authors~(\perc{44})
neither agree nor disagree on whether data snooping is easy to avoid,
emphasizing the importance of clear definitions and recommendations.

\paragraph{Takeaways} We find that all of the pitfalls introduced in
\autoref{sec:pitfalls} are pervasive in security research, affecting
between \perc{17} and \perc{90} of the selected papers. Each paper
suffers from at least three of the pitfalls and only \perc{22} of
instances are accompanied by a discussion in the text. While authors
may have even deliberately omitted a discussion of pitfalls in some
cases, the results of our prevalence analysis overall suggest a lack
of awareness in our community.

Although these findings point to a serious problem in research, we
would like to remark that \textit{all} of the papers analyzed provide
excellent contributions and valuable insights. Our objective here is
not to blame researchers for stepping into pitfalls but to raise
awareness and increase the experimental quality of research on machine
learning in security.

\section{Impact Analysis}
\label{sec:casestudy}

In the previous sections, we have presented pitfalls that are
widespread in the computer security literature.
However, so far it remains unclear how much the individual pitfalls
could affect experimental results and their conclusions.
In this section, we estimate the experimental impact of some of these
pitfalls in popular applications of machine learning in security.  At
the same time, we demonstrate how the recommendations discussed in
\autoref{sec:pitfalls} help in identifying and resolving these problems.
For our discussion, we consider four popular research topics in
computer security:

\begin{itemize}[noitemsep]
  \item\autoref{subsec:androidmalware}: \noemph{mobile malware detection} (\shortsamplingbias, \shortfalsecausality, and \shortinapprmetrics)
  \item\autoref{subsec:vulndiscovery}: \noemph{vulnerability discovery} (\shortlabelinaccuracy, \shortfalsecausality, and \shortinapprbaseline)
  \item\autoref{subsec:codestylo}: \noemph{source code authorship attribution} (\shortsamplingbias and \shortfalsecausality)
  \item\autoref{subsec:nids}: \noemph{network intrusion detection} (\shortinapprbaseline and \shortlabonlyevaluation)
\end{itemize}

\noindent\begin{minipage}{\columnwidth}%
\begin{mdframed}[style=HighlightFrame]
\textbf{Remark.} For this analysis, we consider state-of-the-art 
approaches for each security domain. We remark that the results 
within this section do not
mean to criticize these approaches specifically; we choose them as they
are \emph{representative} of how pitfalls can impact different domains.
Notably, the fact that we have been able to reproduce the
approaches speaks highly of their academic standard.
\end{mdframed}%
\end{minipage}%

\subsection{Mobile Malware Detection}\label{subsec:androidmalware}

The automatic detection of Android malware using machine learning is a
particularly lively area of research. The design and evaluation of such
methods are delicate and may exhibit some of the previously discussed
pitfalls.
In the following, we discuss the effects of
\samplingbias~(\shortsamplingbias),
\falsecausality~(\shortfalsecausality), and
\inapprmetrics~(\shortinapprmetrics) on learning-based detection in
this context.

\paragraph{Dataset collection} A common source of recent mobile data
is the \emph{\az} project~\cite{AllBisKleTra16}, which collects
Android apps from a large variety of sources, including the official
\emph{\gp} store and several Chinese markets.
At the time of writing it includes more than \num{11}~million 
Android applications from \num{18}~different sources. As well as the
samples themselves, it includes meta-information, such as the number
of antivirus detections. Although \az is an excellent source for
obtaining mobile apps, we demonstrate that experiments may suffer from
severe \samplingbias~(\shortsamplingbias) if the peculiarities of the
dataset are not taken into account. Please note that the following
discussion is not limited to the \az data, but is relevant for the
composition of Android datasets in general.

\begin{figure}[!t]
	\centering
	\includegraphics{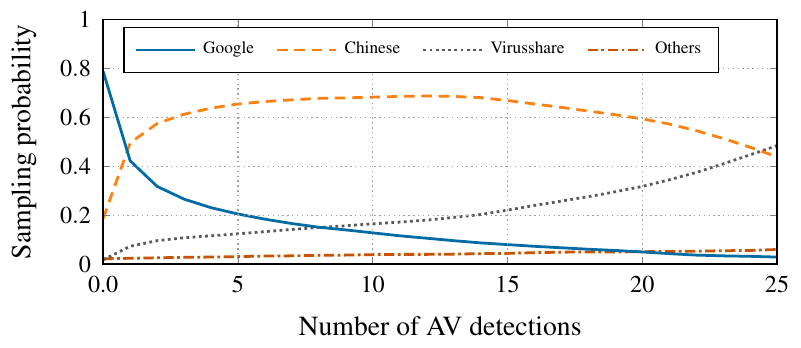}
	\vspace{-8px}
	\caption{The probability of sampling malware from Chinese
          markets is significantly higher than for \gp. This
          can lead to sampling biases in experimental setups for
          Android malware detection.}
	\label{fig:az_sampling_bias}
    \vspace{-1.0em}
\end{figure}

\paragraph{Dataset analysis} In the first step, we analyze the data
distribution of \az by considering the origin of an app and the number
of antivirus detections of an Android app. For our analysis, we
broadly divide the individual markets into four different origins:
\gp, Chinese markets, VirusShare, and all other markets.

\autoref{fig:az_sampling_bias} shows the probability of randomly
sampling from a particular origin depending on the number of
antivirus detections for an app.
For instance, when selecting a sample with no constraints on the
number of detections, the probability of sampling from \gp is
roughly~\perc{80}.
If we consider a threshold of \num{10}~detections, the probability
that we randomly select an app from a Chinese market is~\perc{70}.  
It is very likely that a large fraction of the benign apps in a
dataset are from \gp, while most of the malicious ones originate
from Chinese markets, if we ignore the data distribution.

Note that this sampling bias is not limited to \az. We
identify a similar sampling bias for the \drebin
dataset~\cite{ArpSprHueGasRie14}, which is commonly used to evaluate
the performance of learning-based methods for Android malware
detection~\mbox{\citep[\eg][]{ArpSprHueGasRie14,ZhuDum16,GroPapManBac+17}.}

\newcommand{\wchinese}{$D_1$\xspace}
\newcommand{\wochinese}{$D_2$\xspace}

\paragraph{Experimental setup}
To get a better understanding of this finding,
we conduct experiments using two datasets: For the first
dataset (\wchinese), we merge \num{10000}~benign apps from \gp with
\num{1000}~malicious apps from Chinese markets (\emph{Anzhi} and
\emph{AppChina}). We then create a second dataset (\wochinese)
using the same \num{10000}~benign applications, but combine them with
\num{1000}~malware samples exclusively from \gp. All
malicious apps are detected by at least \num{10}~virus scanners.
Next, we train a linear support vector machine~\citep{FanChaHsiWanLin08}
on these datasets using two feature sets taken from
state-of-the-art classifiers (\drebin~\citep{ArpSprHueGasRie14} and
\opseqs~\citep{McLMarKan17}).

\begin{table}[t]

  \centering
  \caption{Comparison of results for two classifiers when merging
    benign apps from GooglePlay with Chinese malware~(\wchinese) vs.
    sampling solely from GooglePlay~(\wochinese). For both
    classifiers, the detection performance drops significantly when
    considering apps only from GooglePlay. The standard deviation of
    the results ranges between 0--3\%.}
  {  \footnotesize
  \begin{tabular}{
    l
    S[table-format=1.3]
    S[table-format=1.3]
    S[table-format=2.1,table-space-text-post={\perc{}}]
    S[table-format=1.3]
    S[table-format=1.3]
    S[table-format=2.1,table-space-text-post={\perc{}}]
  }
    \toprule
    \multirow[b]{2}{*}[-5pt]{\bfseries Metric} &
    \multicolumn{3}{c}{\bfseries \drebin} &
    \multicolumn{3}{c}{\bfseries \opseqs} \\
    \cmidrule(rl){2-4}\cmidrule(rl){5-7}
     &{ \wchinese }&{ \wochinese }&{ $\Delta $} 
     &{ \wchinese }&{ \wochinese }&{ $\Delta $} \\
    \midrule
    Accuracy &
      0.994 &
      0.980 &
      -1.4\perc{} &
      0.972 &
      0.948 &
      -2.5\perc{} \\
    Precision &
      0.968 &
      0.930 &
      -3.9\perc{} &
      0.822 &
      0.713 &
      -13.3\perc{} \\
    Recall &
      0.964 &
      0.846 &
      -12.2\perc{} &
      0.883 &
      0.734 &
      -16.9\perc{} \\
    F1-Score &
      0.970 &
      0.886 &
      -8.7\perc{} &
      0.851 &
      0.722 &
      -15.2\perc{} \\
      MCC~\citep{Mat75} &
      0.963 &
      0.876 &
      -9.0\perc{} &
      0.836 &
      0.695 &
      -16.9\perc{} \\
    \bottomrule
  \end{tabular}} %
  \label{tab:sampling_bias}
\end{table}

\paragraph{Results}
The recall (true positive rate) for \drebin and \opseqs drops by more
than \perc{10} and \perc{15}, respectively, between the datasets
\wchinese and \wochinese, while the accuracy is only slightly affected
(see \autoref{tab:sampling_bias}). Hence, the choice of the
performance measure is crucial (\shortinapprmetrics).
Interestingly, the URL \emph{play.google.com} turns out to be one of
the five most discriminative features for the benign class, indicating
that the classifier has learned to distinguish the origins of Android
apps, rather than the difference between malware and benign
apps~(\shortfalsecausality). Although our experimental setup
overestimates the classifiers' performance by deliberately ignoring
time dependencies (\shortdatasnooping), we can still clearly observe
the impact of the pitfalls. Note that the effect of temporal snooping
in this setting has been demonstrated in previous
work~\citep{AllTaegBisKleTra15, PenPieJor19}.

\subsection{Vulnerability Discovery}\label{subsec:vulndiscovery}
Vulnerabilities in source code can lead to privilege escalation and 
remote code execution, making them a major threat. 
Since the manual search for vulnerabilities is complex and
time consuming, machine learning-based detection approaches
have been proposed in recent years~\citep{LiZouXuOu+18, YamGolArp+14,
GriGriUza+16}. In what follows, we show that a dataset for vulnerability
detection contains artifacts that occur only in one class
(\shortfalsecausality). We also find that
VulDeePecker~\citep{LiZouXuOu+18}, a neural network to detect
vulnerabilities, uses artifacts for classification and that a simple
linear classifier achieves better results on the same dataset
(\shortinapprbaseline). Finally, we discuss how the preprocessing
steps proposed for VulDeePecker make it impossible to decide whether
some snippets contain vulnerabilities or not (\shortlabelinaccuracy).

\paragraph{Dataset collection}
For our analysis we use the dataset published by \citet{LiZouXuOu+18}, which 
contains source code from the National Vulnerability
Database~\citep{web:NVD} and the SARD project~\citep{web:SARD}. We focus
on vulnerabilities related to buffers (CWE-119) and obtain \num{39757}
source code snippets of which \num{10444}~(\perc{26}) are labeled as
containing a vulnerability.

\begin{table}[t]
  \footnotesize
  \centering
  \caption{Different buffer sizes in the Vulnerability Dataset used
	by \citet{LiZouXuOu+18} with their number of occurrences and
	relative frequency in class \code{0}.}
\label{tab:vuldee-tokens}
  \begin{tabular}{
    S[table-format=2]
    S[table-format=4]
    S[table-format=4,table-space-text-post={~~(\perc{99.9}})]
  }
    \toprule
    \bfseries Buffer size & \multicolumn{2}{c}{\bfseries Occurrences} \\
    \cmidrule(rl){2-3}
    &{ Total }&{ In class \code{0} }\\
    \midrule
         3 &   70 &   53~~(\perc{75.7}) \\
        32 &  116 &  115~~(\perc{99.1}) \\
       100 & 6364 & 4315~~(\perc{67.8}) \\
       128 &   26 &   24~~(\perc{92.3}) \\
      1024 &  100 &   96~~(\perc{96.0}) \\
    \bottomrule
  \end{tabular}%
\end{table}

\paragraph{Dataset analysis} We begin our analysis by classifying a
random subset of code snippets by hand to spot possible artifacts in the
dataset. We find that certain sizes of buffers seem to be present
only in one class throughout the samples considered. To investigate, 
we extract the buffer sizes of \code{char} arrays that are
initialized in the dataset and count the number of occurrences in each
class.
We report the result for class \num{0} (snippets without
vulnerabilities) in \autoref{tab:vuldee-tokens} and observe that certain
buffer sizes occur almost exclusively in this class. If the model relies
on buffer sizes as discriminative features for classification, this would 
be a spurious correlation (\shortfalsecausality).

\begin{figure}[b] \centering
	\begin{subfigure}[t]{0.95\columnwidth}
	\input{tables/casestudy-vuldee-intro-code.tex}
	\vspace{2mm}
	\end{subfigure}
	\begin{subfigure}[t]{0.95\columnwidth} \input{tables/intro-lrp.tex}
	\end{subfigure}
	\caption{Top: Code snippet from the dataset. Bottom: Same code
	snippet after preprocessing steps of VulDeePecker. Coloring
	indicates importance towards classification according to the
	LRP~\citep{BacBinMon+15} method.}
	\label{fig:vuldee-code}
\end{figure}

\paragraph{Experimental setup} We train VulDeePecker~\cite{LiZouXuOu+18}, 
based on a recurrent neural network~\citep{HocSch97}, to classify the
 code snippets automatically. 
To this end, we replace variable names with generic identifiers
(\eg~\code{INT2}) and truncate the snippets to \num{50}~tokens, as
proposed in the paper~\cite{LiZouXuOu+18}. 
An example of this procedure can be seen in
\autoref{fig:vuldee-code} where the original code snippet (top) is
transformed to a generic snippet~(bottom).

We use a linear Support Vector Machine (SVM) with bag-of-words
features based on $n$-grams as a baseline for VulDeePecker. To see
what VulDeePecker has learned we follow the work of
\citet{WarArpWres+20} and use the Layerwise Relevance
Propagation~(LRP) method~\citep{BacBinMon+15} to explain the
predictions and assign each token a \emph{relevance} score that
indicates its importance for the classification.
\autoref{fig:vuldee-code}~(bottom) shows an example for these scores
where blue tokens favor the classification and orange ones oppose it.

\paragraph{Results} To see whether VulDeePecker relies on artifacts, 
we use the relevance values for the entire training set and
extract the ten most important tokens for each code snippet.
Afterwards we extract the tokens that occur most often in this
top-\num{10} selection and report the results in
\autoref{tab:vuldee-artifacts} in descending order of occurrence.

\begin{table}[!t]
  \footnotesize
  \centering
  \caption{The \num{10} most frequent tokens across samples in the dataset.}
  \begin{tabularx}{\linewidth}
    { S[table-format=2]
      Xc
      S[table-format=2]
      Xc
    }
  \toprule
    \bfseries Rank & \bfseries Token & \bfseries Occurrence &
    \bfseries Rank & \bfseries Token & \bfseries Occurrence \\
    \cmidrule{1-3}\cmidrule(l){4-6}
    1   & \code{INT1} & \perc{70.8} &
      6 & \code{char} & \perc{38.8} \\
    2   & \code{(}    & \perc{61.1} &
      7 & \code{]}    & \perc{32.1} \\
    3   & \code{*}    & \perc{47.2} &
      8 & \code{+}    & \perc{31.1} \\
    4   & \code{INT2} & \perc{45.7} &
      9 & \code{VAR0} & \perc{28.7} \\
    5   & \code{INT0} & \perc{38.8} &
     10 & \code{,}    & \perc{26.0} \\
    \bottomrule
  \end{tabularx}%
  \label{tab:vuldee-artifacts}
\end{table}

While the explanations are still hard to interpret for a human we notice
two things: Firstly, tokens such as `\code{(}', `\code{]}', and
`\code{,}' are among the most important features throughout the training
data although they occur frequently in code from both classes as part of
function calls or array initialization. Secondly, there are many generic
\code{INT*} values which frequently correspond to buffer sizes. From this
 we conclude that VulDeePecker is relying on combinations of artifacts
in the dataset and thus suffers from \falsecausality
(\shortfalsecausality).

\begin{table}[b]
    \footnotesize
	\centering
	\caption{Performance of Support Vector Machines and VulDeePecker on
	unseen data. The true-positive rate is determined at \perc{2.9} false positives.}
	\begin{tabular}{
	  l
	  S[table-format=1.1e-1, table-align-exponent=true]
	  S[table-format=1.3]
	  S[table-format=1.3]
	}
		\toprule
		{\bfseries Model        } &
		{\bfseries \# parameters} &
		{\bfseries AUC          } &
		{\bfseries TPR          }\\
		\cmidrule{1-4}
		VulDeePecker
		& 1.2e6
		& 0.984
		& 0.818\\
		SVM
		& 6.6e4
		& 0.986
		& 0.963\\
		AutoSklearn
		& 8.5e5
		& 0.982
		& 0.894\\
		\bottomrule
	\end{tabular}%
	\label{tab:vuldee-baseline}
\end{table}

To further support this finding, we show in
\autoref{tab:vuldee-baseline} the performance of VulDeePecker compared
to an SVM and an ensemble of standard models, such as random forests
and AdaBoost classifiers, trained with the \emph{AutoSklearn}
library~\citep{FeuKleEgg15}. We find that an SVM with \num{3}-grams
yields the best performance with an $18\times$ smaller model.
This is interesting as overlapping but independent substrings
(\mbox{$n$-grams}) are used, rather than the true sequential ordering
of all tokens as for the RNN. Thus, it is likely that VulDeePecker is
not exploiting relations in the sequence, but merely combines special
tokens---an insight that could have been obtained by training a linear
classifier (\shortinapprbaseline). Furthermore, it is noteworthy that
both baselines provide significantly higher true positive rates,
although the AUC-ROC of all approaches only slightly differs.

Finally, we discuss the preprocessing steps proposed by
\citet{LiZouXuOu+18} as seen in the example of
\autoref{fig:vuldee-code}. By truncating the code snippets to a fixed
length of \num{50}, important information is lost. For example, the
value of the variable \code{SRC\_STRING} and thus its length is unknown
to the network. Likewise, the conversion of numbers to \code{INT0} and
\code{INT1} results in the same problem for the \code{data} variable:
after the conversion it is not possible to tell how big the buffer is
and whether the content fits into it or not. Depending on the surrounding
code it can become impossible to say whether buffer overflows appear
or not, leading to cases of \labelinaccuracy~(\shortlabelinaccuracy).

\subsection{Source Code Author Attribution}\label{subsec:codestylo}

The task of identifying the developer based on source code is known as
authorship attribution~\cite{CalHarLiuNar+15}. Programming habits are
characterized by a variety of stylistic patterns, so that
state-of-the-art attribution methods use an expressive set of such
features.
These range from simple layout properties to more unusual habits in the
use of syntax and control flow. In combination with 
\samplingbias~(\shortsamplingbias), this expressiveness may give rise to
\falsecausalities~(\shortfalsecausality) in current attribution methods,
leading to an overestimation of accuracy.

\paragraph{Dataset collection}
Recent approaches have been tested on data from the Google Code Jam
(GCJ) programming competition~\cite{CalHarLiuNar+15, AlsDauHarMan+17,
AbuAbuMoh+18}, where participants solve the same challenges in various
rounds.
An advantage of this dataset is that it ensures a classifier learns to
separate stylistic patterns rather than merely overfitting to different
challenges.
We use the 2017 GCJ dataset~\cite{QuiMaiRie19}, which consists of
\num{1632}~\CC~files from \num{204}~authors solving the same
eight~challenges.

\begin{figure}[t]
	\centering
	\includegraphics{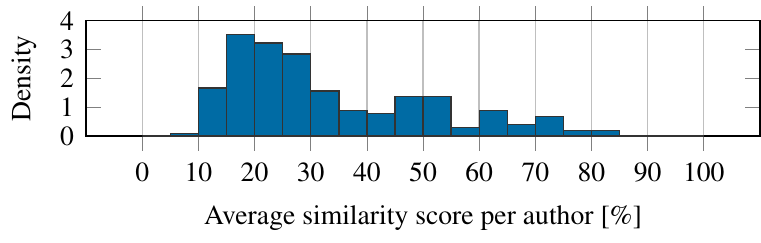}
	\caption{Shared source code over all files per author. A majority
	tend to copy code snippets across challenges, leading to learned
	artifacts.}
	\label{fig:codestylo_dataset}
        \vspace{0.2em}
\end{figure}

\paragraph{Dataset analysis}
We start with an analysis of the average similarity score between all
files of each respective programmer, where the score is computed by
\emph{difflib's SequenceMatcher}~\citep{web:difflib}.
\autoref{fig:codestylo_dataset} shows that most participants copy
code across the challenges, that is, they reuse personalized coding
\emph{templates}.  Understandably, this results from the nature of 
the competition, where participants are encouraged to solve challenges
quickly.
These templates are often \emph{not} used to solve the current
challenges but are only present in case they might be needed.
As this deviates from real-world settings, we identify a
\samplingbias in the dataset.

Current feature sets for authorship attribution include these templates,
such that models are learned that strongly focus on them as highly
discriminative patterns.
However, this unused duplicate code leads to features that represent
artifacts rather than coding style which are \falsecausalities.

\paragraph{Experimental setup}
Our evaluation on the impact of both pitfalls builds on the
attribution methods by \mbox{\citet{AbuAbuMoh+18}} and
\citet{CalHarLiuNar+15}. Both represent the state of the art regarding
performance and comprehensiveness of features. 

We implement a linter tool on top of Clang, an open-source C/\CC
front-end for the LLVM compiler framework, to remove unused code that is
mostly present due to the templates. Based on this, we design the
following three experiments: First, we train and test a classifier on
the unprocessed dataset~($T_b$) as a baseline. Second, we remove unused
code from the respective test sets~($T_1$), which allows us to test how
much the learning methods focus on unused template code. Finally, we
remove unused code from the training set and re-train the
classifier~($T_2$).

\begin{figure}[t]
	\centering
	\includegraphics{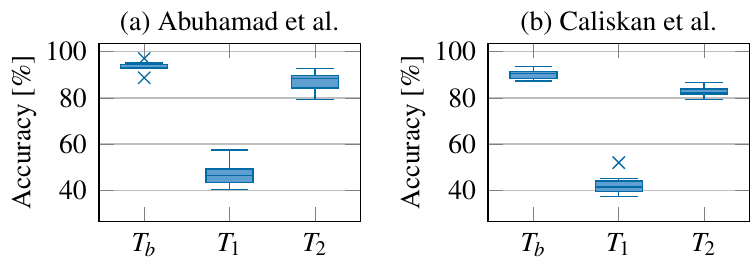}
	\vspace{-0.3em}
	\caption{Accuracy of authorship attribution after considering
	artifacts. The accuracy drops by \perc{48} if unused code is removed
	from the test set ($T_1$); After retraining ($T_2$), the
	average accuracy still drops by \perc{6} and \perc{7}.}
        \label{fig:codestylo_accuracy}
\end{figure}

\paragraph{Results}
\autoref{fig:codestylo_accuracy} presents the accuracy for both
attribution methods on the different experiments. Artifacts have a
substantial impact on the attribution accuracy.
If we remove unused code from the test set~($T_1$), the accuracy drops
by~\perc{48} for the two approaches. This shows both systems focus 
considerably on the unused template code.
After retraining~($T_2$), the average accuracy drops by~\perc{6} and
\perc{7} for the methods of \citet{AbuAbuMoh+18} and
\mbox{\citet{CalHarLiuNar+15}}, demonstrating the reliance on
artifacts for the attribution performance.

Overall, our experiments show that the impact of \samplingbias and
\falsecausality has been underestimated and reduces the accuracy
considerably.
At the same time, our results are encouraging. After accounting for 
artifacts, both attribution methods select features that allow for a 
more reliable identification. We make the sanitized dataset publicly
available to foster further research in this direction.

\subsection{Network Intrusion Detection}
\label{subsec:nids}

Detecting network intrusions is one of the oldest problems in
security~\citep{denning1986ids} and it comes at no surprise that
detection of anomalous network traffic relies heavily on learning-based
approaches~\cite{LeeSto98, ChaBanKum09, LiXiaZhangYanAiDai12,
MirDoiEloSha18}.
However, challenges in collecting real attack data~\citep{EngRimJoo21}
has often led researchers to generate synthetic data for
\labonlyevaluations~(\shortlabonlyevaluation).
Here, we demonstrate how this data is often insufficient for justifying
the use of complex models (\eg~neural networks) and how using a simpler
model as a baseline would have brought these shortcomings to
light~(\shortinapprbaseline).

\paragraph{Dataset collection} We consider the dataset released
by~\citet{MirDoiEloSha18}, which contains a capture of Internet of
Things~(IoT) network traffic simulating the initial activation and
propagation of the Mirai botnet malware. The packet capture covers
\num{119}~minutes of traffic on a Wi-Fi network with three PCs and nine
IoT~devices. %

\begin{figure}[!t]
    \centering
    \noindent
\includegraphics[width=1\columnwidth]%
	{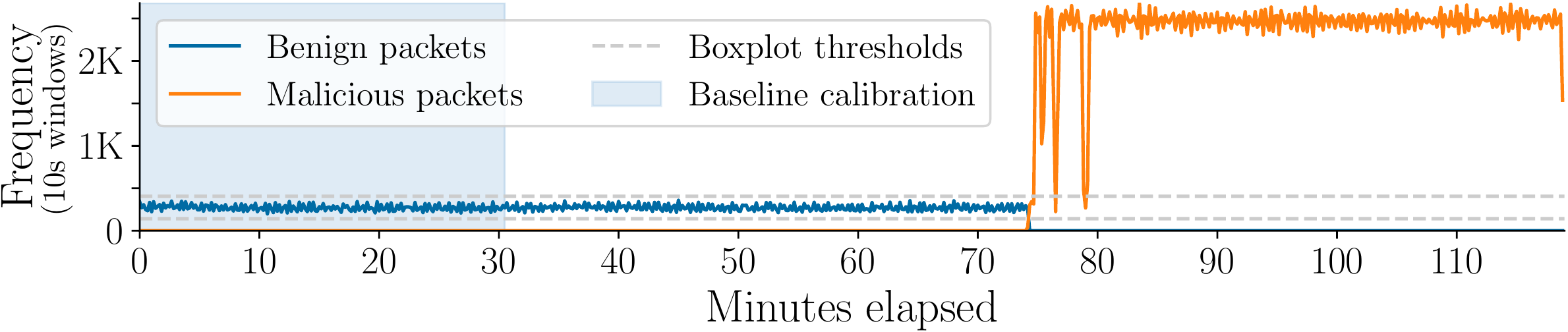}
	\caption{Frequency of benign vs malicious packets in the Mirai 
	dataset~\cite{MirDoiEloSha18}. The Gray dashed lines show the 
	thresholds that define normal traffic calculated using the simple baseline
    (\emph{boxplot method}~\citep{tukey1977}). The span of clean
    data used for calibration is highlighted by the light blue shaded
    area.}
    \label{fig:kitsune-mirai-freq}
\end{figure}

\paragraph{Dataset analysis} First, we analyze the transmission volume
of the captured network traffic. \autoref{fig:kitsune-mirai-freq} shows
the frequency of benign and malicious packets across the capture,
divided into bins of \num{10}~seconds.
This reveals a strong signal in the packet frequency, which is highly
indicative of an ongoing attack. Moreover, all benign activity seems to
halt as the attack commences, after \num{74}~minutes, despite
the number of devices on the network. This suggests that individual
observations may have been merged and could further result in
the system benefiting from \falsecausality~(\shortfalsecausality).

\paragraph{Experimental setup} To illustrate how severe these pitfalls are, we consider \kitsune~\cite{MirDoiEloSha18}, a
state-of-the-art deep learning-based intrusion detector built on an
ensemble of autoencoders.
For each packet, \num{115}~features are extracted that are input to
\num{12}~autoencoders, which themselves feed to another, final
autoencoder operating as the anomaly detector.

As a simple baseline to compare against \kitsune, we choose the
\emph{boxplot method}~\citep{tukey1977}, a common approach for
identifying outliers.
We process the packets using a \num{10}-second sliding window and use
the packet frequency per window as the sole feature. Next, we derive a
lower and upper threshold from the clean calibration distribution:
    $\tau_{low} = Q_1 - 1.5 \cdot \mathrm{IQR}$ and $\tau_{high} = Q_3 +
    1.5 \cdot \mathrm{IQR}$.
During testing, packets are marked as benign if the sliding window's
packet frequency is between $\tau_{low}$ and $\tau_{high}$, and
malicious otherwise. In~\autoref{fig:kitsune-mirai-freq}, these
thresholds are shown by the dashed gray lines.

\begin{table}[!t]
    \footnotesize
    \centering
    \caption{Comparing \kitsune~\cite{MirDoiEloSha18}, an autoencoder
    ensemble NIDS, against a simple baseline, boxplot
    method~\cite{tukey1977}, for detecting a Mirai infection.}
    \begin{tabular}{
      p{3cm}
      S[table-format=1.3]
      S[table-format=1.3]
      S[table-format=1.3]
     }
        \toprule
        {\bfseries Detector } &
        {\bfseries AUC      } &
        {\bfseries TPR      } &
        {\bfseries TPR      }\\
        &&\scriptsize (FPR at \num{0.001})
        & \scriptsize (FPR at \num{0.000})\\
        \midrule
        \kitsune~\citep{MirDoiEloSha18}
        & 0.968
        & 0.882
        & 0.873\\
        Simple Baseline~\citep{tukey1977}
        & 0.998
        & 0.996
        & 0.996\\
        \bottomrule
    \end{tabular}%
    \label{tab:nids-baseline}
\end{table}

\paragraph{Results} The classification performance of the autoencoder
ensemble compared to the boxplot method is shown
in~\autoref{tab:nids-baseline}. While the two approaches perform
similarly in terms of ROC AUC, the simple boxplot method outperforms the
autoencoder ensemble at low false-positive rates~(FPR).
As well as its superior performance, the boxplot
method is exceedingly lightweight compared to the feature extraction and
test procedures of the ensemble. This is especially relevant as the
ensemble is designed to operate on resource-constrained devices
with low latency (\eg~IoT~devices).

Note this experiment does not intend to show that the boxplot
method can detect an instance of Mirai operating in the wild, nor that
\kitsune is incapable of detecting other attacks, but to
demonstrate that an experiment without an
\apprbaseline~(\shortinapprbaseline) \textit{is insufficient to justify the
complexity and overhead of the ensemble}.
The success of the boxplot method also shows how simple methods can 
reveal issues with data generated for 
\labonlyevaluations~(\shortlabonlyevaluation). In the Mirai dataset  
the infection is overly conspicuous; an attack 
in the wild would likely be represented by a tiny proportion of 
network traffic.

\subsection{Takeaways} 
The four case studies clearly demonstrate the impact of the considered
pitfalls across four distinct security scenarios. Our findings show
that subtle errors in the design and experimental setup of an approach
can result in misleading or erroneous results. Despite the overall
valuable contributions of the research, the
frequency and severity of pitfalls identified in top papers clearly 
indicate that
significantly more awareness is needed. Additionally, we show how
pitfalls apply across multiple domains, indicating a general problem that
cannot be attributed to only one of the security areas.

\section{Limitations and Threats to Validity}
\label{sec:limitations}
The preceding identification and analysis of common pitfalls in the
security literature has been carried out with utmost care. However,
there are some limitations that are naturally inherent to this kind of
work. Even though these do not affect the overall conclusion of our
analysis, we discuss them in the following for the sake of
completeness.

\paragraph{Pitfalls}
Although some pitfalls may seem obvious at first, our prevalence 
analysis indicates the opposite. This
lack of awareness obstructs progress, and it will persist until 
addressed by the community. Furthermore, we cannot cover all ten
pitfalls in detail, as our focus is on a comprehensive overview. 
Finally, some pitfalls cannot always be prevented, such as 
sampling bias, label inaccuracy, or lab-only settings. For example, 
it is likely not possible to test an attack in a real environment 
due to ethical considerations. In such cases, simulation is the only 
option. 
As outlined in \autoref{sec:pitfalls}, corrective measures may even 
be an open problem, yet awareness of pitfalls is a first step towards 
amending experimental practices and ultimately devising novel methods 
for mitigating them.

\paragraph{Prevalence analysis} For the prevalence
analysis, we skimmed all papers of top security conferences in the
last \num{10}~years and identified \num{30}~papers that use machine learning
prominently (\eg mentioned in the abstract or introduction). Even
though this selection process is not entirely free from bias, the
identified pitfalls are typical for this research branch and
the respective papers are often highly cited. %

Moreover, a pitfall is only counted if its
presence is clear from the text or the associated artifacts, such as
code or data. Otherwise, we decide in favor of the paper and consider a 
pitfall as not present. Despite this conservative assignment, 
our analysis underlines the prevalence of pitfalls.

\paragraph{Impact analysis} Four exemplary research
works are chosen from security areas in which the authors of this paper
have also published research. This biased selection, however, should be 
acceptable, as we intend to empirically demonstrate how
pitfalls can affect experimental results.

\section{Related Work\resp{Fabio (+Daniel)}}
\label{sec:related_work}

Our study is the first to \emph{systematically} and
\emph{comprehensively} explore pitfalls when applying machine learning
to security. It complements a series of research on improving
experimental evaluations in general. In the following, we briefly review
this related work and point out key differences.

\paragraph{Security studies}
Over the last two decades, there have been several studies on
improving experiments in specific security domains. For example,
\citet{Axe00}, \citet{McH00}, and \citet{CarBarSea06} investigate
issues with the evaluation of intrusion detection systems, covering
special cases of sampling bias~(\shortsamplingbias), the base rate
fallacy~(\shortbaseratefallacy), and inappropriate performance
measures~(\shortinapprmetrics). \citet{sommer2010outside} extend this work 
and specifically focus on the application of machine learning for network
intrusion detection. They identify further issues, such as semantic
gaps with anomaly detection~(\shortfalsecausality) and unrealistic
evaluation baselines~(\shortinapprbaseline).

In a similar strain of research, \citet{RosDieGriKrePaxPohBosSte12}
derive guidelines for conducting experiments with malware. Although
this study does not investigate machine learning explicitly, it points
to experimental problems related to some of the issues discussed in
this paper. The study is expanded upon by a series of work examining
variants of sampling bias in malware
analysis~(\shortsamplingbias), such as temporally inconsistent data
splits and labels~\citep[\eg][]{PenPieJor19, AllTaegBisKleTra15,
  MilKanTscAfrBacFai16, ZhuShiYan+20} as well as unrealistic
goodware-to-malware ratios~\citep[\eg][]{Allix:Empirical,
  PenPieJor19}. \citet{packware:ndss20} study the limits of static
features for malware classification in the presence of packed samples.

\citet{DasWerAntPolMon19} show that security
defenses relying on hardware performance counters are ineffective
in realistic settings~(\shortlabonlyevaluation). Similarly, for
privacy-preserving machine learning, \citet{OyaTroGon19} find that
most location privacy approaches fail when applied to real-world
distributions~(\shortlabonlyevaluation). For authentication,
\citet{SugLiuLeaLin19} propose appropriate measures to evaluate
learning-based authentication systems~(\shortinapprmetrics), and 
finally, for system security, \citet{KouHeiAndBosGiu19} point to
frequent benchmarking flaws~(\shortsamplingbias, \shortinapprbaseline,
and \shortinapprmetrics).

Our study builds on this research but provides an orthogonal and
comprehensive view of the problem. Instead of focusing on specific
domains, we are the first to \emph{generally} explore pitfalls and
recommendations when applying machine learning in computer
security. Hence, our work is not limited to certain
problems but applicable to all security domains.

\paragraph{Adversarial learning studies}
Another branch of research has focused on attacking and defending
learning algorithms~\cite{PapMcSin+18, biggio2018wild,
  privacyml2020overview}. While a number of powerful attacks have
emerged from this research such as evasion, poisoning, and inference
attacks, the corresponding defenses have often suffered from limited
robustness~\citep{athalye2018obfuscated}. To counteract this
imbalance, \citet{CarAthPapBreRauTsiGooMadKur19} identify several
pitfalls that affect the evaluation of defenses and discuss
recommendations on how to avoid them. In a similar vein,
\citet{Biggio:Security} propose a framework for security evaluations
of pattern classifiers under attack. Both works are closely related to
pitfall~\shortadversarialsetting and provide valuable hints for
evaluating the robustness of defenses. However, while we also argue
that smart and adaptive adversaries must always be considered when
proposing learning-based solutions in security, our study is more
general.

\paragraph{Machine learning studies}
Finally, a notable body of work has explored recommendations for the
general use of machine learning. This research includes studies on
different forms of sampling bias and dataset
shift~\mbox{\citep{Torralba:Unbiased, Moreno:Unifying, arxiv:SurGut19}}
as well as on the general implications of biased parameter
selection~\citep{HeaHolLan15}, data snooping \citep{KomMae18}, and
inappropriate evaluation methods~\citep{ForSch10, davis2006relationship,
  hand2009measuring}.  An intuitive overview of issues in
applied statistics is provided by \citet{Rei15}.

Our work builds on this analysis; however, we focus exclusively on 
the impact of pitfalls prevalent in security.
Consequently, our study and its recommendations are tailored to the
needs of the security community, and aim to push forward the state of
the art in learning-based security systems. 

\section{Conclusion}
\label{sec:conclusion}
We identify and systematically assess ten~subtle pitfalls in the use
of machine learning in security. These issues can
affect the validity of research and lead to overestimating the
performance of security systems.
We find that these pitfalls are prevalent in security research, and 
demonstrate the impact of these pitfalls in  different security 
applications. To support researchers in avoiding them, we provide 
recommendations that are applicable to all security domains, from 
intrusion and malware detection to vulnerability discovery.

Ultimately, we strive to improve the scientific quality of
empirical work on machine learning in security. 
A decade after the seminal study of
\citet{sommer2010outside}, we again encourage the community to reach
\emph{outside the closed world} and explore the challenges and chances
of embedding machine learning in real-world security systems.

\section*{Additional material}

For interested readers, we provide supplementary material for the
paper at \url{http://dodo-mlsec.org}. 

\section*{Acknowledgements}

The authors wish to thank the anonymous reviewers for their insightful
and constructive comments on this paper. Also, we thank Melanie
Volkamer and Sascha Fahl for their valuable feedback on the study
design. Furthermore, we like to thank Christopher J. Anders for his
helpful suggestions on a previous version of the paper.  The authors
gratefully acknowledge funding from the German Federal Ministry of
Education and Research (BMBF) as BIFOLD -- Berlin Institute for the
Foundations of Learning and Data (ref.  01IS18025A and ref
01IS18037A), by the Helmholtz Association (HGF) within topic ``46.23
Engineering Secure Systems'', and by the Deutsche
Forschungsgemeinschaft (DFG, German Research Foundation) under
Germany's Excellence Strategy EXC 2092 CASA-390781972, and the
projects 456292433;~456292463;~393063728. Moreover, we acknowledge
that this research has been partially sponsored by the UK EP/P009301/1
EPSRC research grant.

{\scriptsize
\setlength{\bibsep}{4pt}

}
\appendix

\section{Appendix: Pitfalls}
\label{sec:appendix_pitfalls}

As a supplement to this paper, we provide further details on the
identified pitfalls and our recommendations in this section.

\paragraph{Label inaccuracy}
Noisy labels are a common problem in machine learning and a  
source of bias. 
In contrast to sampling bias, however, there exist different, practical
methods for mitigating label noise~\citep[\eg][]{NorJiaChu21, 
XuLiDen21}. 
To demonstrate this mitigation, we employ the readily available method
by~\citet{NorJiaChu21} that cleans noisy instances in the training data.
We follow the setup from \citet{XuLiDen21} and randomly flip \perc{9.7} of
the labels in the \drebin training dataset. We then train an SVM on
three datasets: the correctly-labelled dataset, its variant with noisy
labels, and the cleansed dataset.

\autoref{tab:appendix-label-inaccuracy} shows the F1-score, Precision,
and Recall for the three datasets. Due to label noise, the F1-score
drops from \num{0.95}~to~\num{0.73} on the second dataset. Yet, it
increases to~\num{0.93} once data cleansing is applied. This result is
comparable to the method by \citet{XuLiDen21} who report an F1-score
of~\num{0.84} after repairing labels. In addition, we check the
detection performance of noisy labels. \perc{84}~of the flipped labels
are correctly detected, while only \perc{0.2} of the original labels are
falsely flagged as incorrect.
Our experiment indicates that available methods for reducing label noise
can provide sufficient quality to mitigate \labelinaccuracy in practice.

\begin{table}[ht]
\centering
\caption{Performance with label noise in the \drebin dataset}
{  \footnotesize
\begin{tabular}{
  l
  S[table-format=1.3]
  S[table-format=1.3]
  S[table-format=1.3]
}
	\toprule
	{\bfseries Scenario  } &
	{\bfseries F1-Score  } &
	{\bfseries Precision } &
	{\bfseries Recall    }
	\\
	\midrule
Correctly-labelled dataset &     0.955 &      0.900 &   0.928 \\
Noisy dataset              &     0.727 &      0.340 &   0.942 \\
Cleansed dataset           &     0.933 &      0.889 &   0.855 \\
	\bottomrule
\end{tabular}} %
\label{tab:appendix-label-inaccuracy}
\end{table}

\paragraph{Data snooping}
As discussed in \autoref{sec:pitfalls}, there exist several variants of
data snooping where information that is not available in practice is
unintentionally used in the learning process.
\autoref{tab:appendix-snooping} provides a list of common types for
test, temporal and spatial snooping to better illustrate these cases. We
recommend using this table as a starting point when vetting a
machine-learning workflow for the presence of data snooping.

\paragraph{Spurious correlations} Various extraneous
factors, including sampling bias and confounding
bias~\citep{BarTiaPea14,PeaMadJew16}, can introduce spurious
correlations. In the case of confounding bias, a so-called confounder
is present that coincidentally correlates with the task to solve.
Depending on the used features, the confounder introduces artifacts
that lead to false associations. In the case of sampling bias, 
the correlations result from differences between the sampled data and the true
underlying data distribution.

Spurious correlations are challenging to identify, as they depend on the
application domain and the concrete objective of the learning-based
system. In one setting a correlation might be a valid signal, whereas in 
another it spuriously creates an artificial shortcut leading to
over-estimated results. Consequently, we recommend systematically
analyzing possible factors that can introduce these correlations. In some
cases, it is then possible to explicitly control for unwanted extraneous
factors that introduce spurious correlations, thus eliminating their
impact on the experimental outcome.
In other disciplines, different techniques have been proposed to
achieve this goal~\citep[\eg][]{Heck79, Zad04, BarTiaPea14, LiuZie14,
  OvaAhsZha20}, which, however, often build on information not
available to security practitioners. For instance, several methods
\citep[\eg][]{Heck79,Zad04} can correct sampling bias if the selection
probability for each observation is known or can be estimated. In
security research this is rarely the case.

As a remedy, we encourage the community to continuously check for
extraneous factors that affect the performance of learning-based
systems in experiments. However, this is a non-trivial task, as the factors
contributing to the correlations are highly domain-specific. As
recommended in \autoref{sec:pitfalls}, explanation techniques for
machine learning can be a powerful tool in this setting to enable 
tracing predictions back to individual features, thereby exposing the
learned correlations.

\begin{table}[b]
	\centering
	\caption{Detection performance on data from different origins}
	{  \footnotesize
		\begin{tabular}{
				l
				S[table-format=1.3]  
				S[table-format=1.3]
				S[table-format=1.3]
			}
			\toprule
			{\bfseries Origin    } &
			{\bfseries F1-Score  } &
			{\bfseries Precision } &
			{\bfseries Recall    }
			\\
			\midrule
			GooglePlay &     0.879 &      0.914 &   0.846 \\
			Anzhi      &     0.838 &      0.881 &   0.801 \\
			AppChina   &     0.807 &      0.858 &   0.762 \\
			\midrule
			AndroZoo   &     0.885 &      0.922 &   0.852 \\
			\bottomrule
	\end{tabular}} %
	\label{tab:appendix-origin-analysis}
\end{table}

\newcommand{\tabspac}{\addlinespace[4pt]}
\begin{table*}[htbp]
	\centering
	\footnotesize
	\caption{Overview of data snooping groups and types}
	\label{tab:appendix-snooping}

	\begin{tabularx}{\textwidth}{p{.13\textwidth}p{.15\textwidth}X}
	\toprule
	\head{Group} & \head{Types} & \head{Description}
	\\
	\midrule
	\addlinespace[5pt]
	Test Snooping & Preparatory work & 
        If the test set is used for any experiments except for the
        evaluation of the final model, the learning setup benefits
        from additional knowledge that would not be available in
        practice.  This includes steps to find features or limit the
        number of features through feature selection on the entire
        dataset.
	\\ \tabspac
	& K-fold cross-validation & 
	Another type of snooping occurs if researchers tune the 
	hyperparameters by using k-fold cross-validation with the final 
	test set for evaluation, and report these results. 
	\\ \tabspac
	& Normalization & 
	Normalization factors, such as tf-idf, are computed on the 
	complete dataset, i.e., before splitting the dataset into 
	training and test set.
	\\ \tabspac
	& Embeddings & Similarly,
	embeddings for deep neural networks are derived from the complete 
	dataset, instead of just using the training data. 
	\\
	\midrule
	\addlinespace[5pt]
	Temporal Snooping & Time dependency & 
        Time dependencies within the data are not considered, so that
        samples are detected with features that would not be available
        at training time in a realistic setting (\eg features of new
        malware variants~\cite{PenPieJor19}). 
	\\ \tabspac
	& Aging datasets & 
        The usage of well-known datasets from prior work can also
        introduce a bias. Researchers may implicitly incorporate prior
        knowledge by using previous insights from these publicly
        available datasets.
	\\
	\midrule
	\addlinespace[5pt]
	Selective Snooping & Cherry-picking & Data is cleaned based on 
	information that is usually not available in practice. For instance,
	applications are filtered out that are not detected by a
	sufficiently large number of AV scanners. 
	\\ \tabspac
	& Survivorship bias & A group of samples is already filtered out. 
	This bias overlaps with \samplingbias (\shortsamplingbias).
	For example, using only applications, 
	which a dynamic analysis system can successfully process and 
	removing all others from the dataset, also introduces a 
	survivorship bias.   
	\\
	\bottomrule
	\end{tabularx}

\end{table*}

\paragraph{Sampling bias}
Often it is extremely difficult to acquire representative data and thus
some bias is unavoidable. As an example of how to tackle this problem, we
investigate this pitfall for Android malware detection. In particular,
we control for one source of sampling bias to prevent our classifier from
picking up on spurious correlations, rather than detecting malware.
To this end, we construct individual datasets that exclusively contain
only apps from one specific market each, instead of training on the
overall, large dataset of Android apps.
This ensures that the classifier learns to detect malware instead of
capturing differences between the markets. For this experiment, we use
the three largest markets in AndroZoo (GooglePlay, Anzhi, and AppChina)
with \num{10000}~benign and \num{1000}~malicious apps each, and train
\drebin on all datasets.

The results are depicted in Table~\ref{tab:appendix-origin-analysis}.
The detection performance varies across the datasets, with an F1-score
ranging from \num{0.807} to \num{0.879}. However, if we ignore the
origin of the apps and randomly sample from the complete AndroZoo
dataset, we obtain the best F1-score of \num{0.885}, indicating a
clear sampling bias. This simple experiment demonstrates how 
controlling for a source of bias can help to better estimate the 
performance of a malware detector. While the example is simple 
and specific to Android malware, it is easily transferable to 
other sources and scenarios.

\section{Appendix: Prevalence Analysis}
\label{sec:appendix-survey}

Here, we provide additional details on the author survey discussed in
\autoref{sec:prevalence}. Note that the supplementary material
contains further information regarding the chosen papers.

\begin{figure}[h]
  \centering 
  \includegraphics{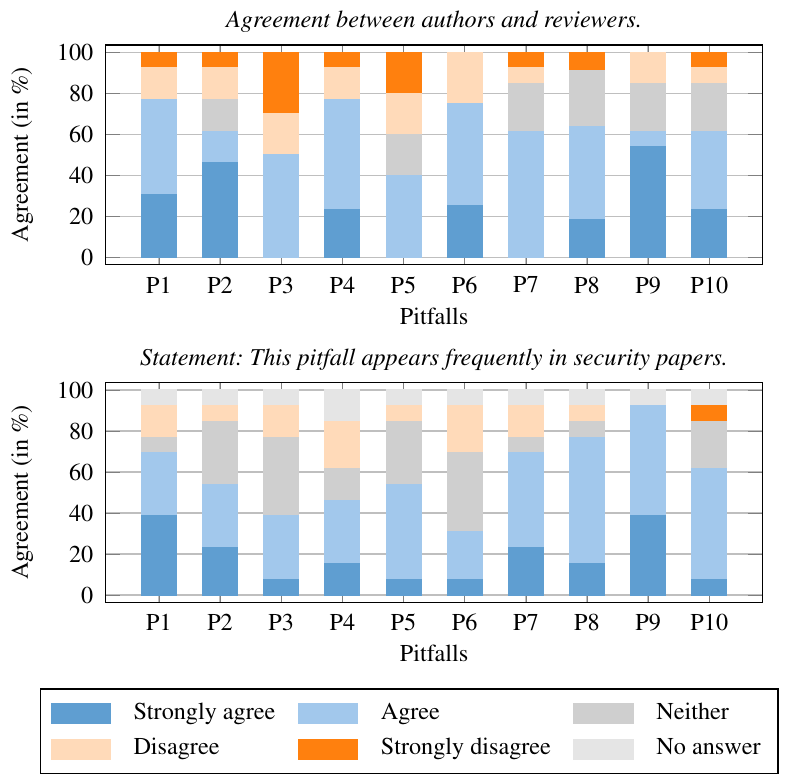}
  \caption{Survey results regarding the different pitfalls.}
  \label{fig:survey-results}
\end{figure}

\paragraph{Details of author survey} In addition to the discussion of
the survey conducted in \autoref{sec:prevalence},
\autoref{fig:survey-results} provides an overview of the authors'
responses grouped by pitfall. Each bar indicates the agreement of the
authors, with colors ranging from warm (strongly disagree) to cold
(strongly agree).

\paragraph{Data collection and ethics} Our institution does not require
a formal IRB process for the survey conducted in this work. However,
we contacted the ethical review board of our institution and achieved
approval from its chair for conducting the survey. Moreover, we
designed the survey in accordance with the General Data Protection
Regulation of the EU, minimizing and anonymizing data where
possible. All authors approved to a consent form that informed them
about the purpose of the study, the data we collect, and included 
an e-mail address to contact us in case of questions.

\end{document}